\documentstyle[12pt,epsfig]{article}
\newcommand{\av}[1]{\mbox{$ \langle #1 \rangle $}}
\newcommand{\lsim}{\raisebox{-0.5mm}{$\stackrel{<}{\scriptstyle{\sim}}$}}
\newcommand{\gsim}{\raisebox{-0.5mm}{$\stackrel{>}{\scriptstyle{\sim}}$}}
\newcommand{\gsp}{\mbox{$\gamma^* p$}}
\newcommand{\PO}{{\rm l \! P }}
\newcommand{\rh}{\mbox{$\varrho$}}
\newcommand{\rhz}{\mbox{$\rh^0$}}
\newcommand{\ph}{\mbox{$\phi$}}
\newcommand{\om}{\mbox{$\omega$}}
\newcommand{\jpsi}{\mbox{$J/\psi$}}
\newcommand{\Qsq}{\mbox{$Q^2$}}
\newcommand{\qsq}{\mbox{$Q^2$}}
\newcommand{\W}{\mbox{$W$}}
\newcommand{\x}{\mbox{$x$}}
\newcommand{\y}{\mbox{$y$}}
\newcommand{\ttr}{\mbox{$t$}}
\newcommand{\s}{\mbox{$s$}}
\newcommand{\R}{\mbox{$R$}}
\newcommand{\bslope}{\mbox{$b$}}
\newcommand{\eclmax}{\mbox{$E_{max}$}}
\newcommand{\mpipi}{\mbox{$m_{\pi^+\pi^-}$}}
\newcommand{\mll}{\mbox{$m_{l^+l^-}$}}
\newcommand{\mpp}{\mbox{$m_{\pi\pi}$}}       
\newcommand{\mppsq}{\mbox{$m_{\pi\pi}^2$}}       
\newcommand{\mpi}{\mbox{$m_{\pi}$}}          
\newcommand{\mrho}{\mbox{$m_{\rho}$}}        
\newcommand{\mrhosq}{\mbox{$m_{\rho}^2$}}        
\newcommand{\Gmpp}{\mbox{$\Gamma (\mpp)$}}    
\newcommand{\Gmppsq}{\mbox{$\Gamma^2(\mpp)$}}    
\newcommand{\Grho}{\mbox{$\Gamma_{\rho}$}}   
\newcommand{\gp}{\mbox{$\gamma p$}}
\newcommand{\modt}{\mbox{$|t|$}}
\newcommand{\eminpz}{\mbox{$E-p_z$}}
\newcommand{\costhst}{\mbox{$\cos\theta^*$}}
\newcommand{\rzzzz}{\mbox{$r_{00}^{04}$}}

\newcommand{\gev}{\mbox{\rm GeV}}
\newcommand{\gevc}{\mbox{\rm GeV/c}}

\newcommand{\GeVsq}{\mbox{${\rm GeV}^2$}}
\newcommand{\gevsq}{\mbox{${\rm GeV}^2$}}

\newcommand{\gevcsq}{\mbox{${\rm GeV/c^2}$}}
\newcommand{\mevcsq}{\mbox{${\rm MeV/c^2}$}}

\newcommand{\gevsqm}{\mbox{${\rm GeV}^{-2}$}}

\newcommand{\pbinv}{\mbox{${\rm pb^{-1}}$}}
\def\ar#1#2#3   {{\em Ann. Rev. Nucl. Part. Sci.} {\bf#1} (#2) #3}
\def\err#1#2#3  {{\it Erratum} {\bf#1} (#2) #3}
\def\ib#1#2#3   {{\it ibid.} {\bf#1} (#2) #3}
\def\ijmp#1#2#3 {{\em Int. J. Mod. Phys.} {\bf#1} (#2) #3}
\def\jetp#1#2#3 {{\em JETP Lett.} {\bf#1} (#2) #3}
\def\mpl#1#2#3  {{\em Mod. Phys. Lett.} {\bf#1} (#2) #3}
\def\nim#1#2#3  {{\em Nucl. Instr. Meth.} {\bf#1} (#2) #3}
\def\nc#1#2#3   {{\em Nuovo Cim.} {\bf#1} (#2) #3}
\def\np#1#2#3   {{\em Nucl. Phys.} {\bf#1} (#2) #3}
\def\pl#1#2#3   {{\em Phys. Lett.} {\bf#1} (#2) #3}
\def\prep#1#2#3 {{\em Phys. Rep.} {\bf#1} (#2) #3}
\def\prev#1#2#3 {{\em Phys. Rev.} {\bf#1} (#2) #3}
\def\prl#1#2#3  {{\em Phys. Rev. Lett.} {\bf#1} (#2) #3}
\def\ptp#1#2#3  {{\em Prog. Th. Phys.} {\bf#1} (#2) #3}
\def\rmp#1#2#3  {{\em Rev. Mod. Phys.} {\bf#1} (#2) #3}
\def\rpp#1#2#3  {{\em Rep. Prog. Phys.} {\bf#1} (#2) #3}
\def\sjnp#1#2#3 {{\em Sov. J. Nucl. Phys.} {\bf#1} (#2) #3}
\def\spj#1#2#3  {{\em Sov. Phys. JEPT} {\bf#1} (#2) #3}
\def\zp#1#2#3   {{\em Zeit. Phys.} {\bf#1} (#2) #3}
\begin{document}
\begin{titlepage}
\begin{flushleft}
{\tt DESY 96-023 \hfill ISSN 0418-nnnn}\\
{\tt February 1996}\\
\vspace{1cm}

\end{flushleft}
\vspace*{4.cm}
\begin{center}
\begin{Large}
\boldmath
\bf{    Elastic Electroproduction of \rh\ and \jpsi\ Mesons \\
at large \qsq\ at HERA\\}
\unboldmath 

\vspace*{2.cm}
H1 Collaboration \\
\end{Large}

\vspace*{1cm}

\end{center}

\vspace*{1cm}


\vspace*{2.cm}
\begin{abstract}
\noindent
The total cross sections for the elastic electroproduction of \rh\ and
\jpsi\ mesons for \qsq $>$ 8 \gevsq\ and $\langle W \rangle \simeq 90$ \gevcsq\ 
are measured at HERA with the H1 detector. The measurements are for an
integrated electron$-$proton luminosity of $\simeq$~3~pb$^{-1}$. The
dependences of the total virtual photon$-$proton (\gsp) cross
sections on $Q^2$, $W$ and the momentum transfer squared to the proton ($t$), 
and, for the $\rho$, the dependence on the polar decay angle ($\cos \theta^*$), are 
presented. The \jpsi\ : \rh\ cross section ratio is determined. 
The results are discussed in the light of theoretical models 
and of the interplay of hard and soft physics processes.

\end{abstract}
\end{titlepage}

\noindent
 S.~Aid$^{14}$,                   
 V.~Andreev$^{26}$,               
 B.~Andrieu$^{29}$,               
 R.-D.~Appuhn$^{12}$,             
 M.~Arpagaus$^{37}$,              
 A.~Babaev$^{25}$,                
 J.~B\"ahr$^{36}$,                
 J.~B\'an$^{18}$,                 
 Y.~Ban$^{28}$,                   
 P.~Baranov$^{26}$
% ...
 
%
%
%
\noindent
 $\:^1$ I. Physikalisches Institut der RWTH, Aachen, Germany$^ a$ \\
% ...
$ ^{12}$ DESY, Hamburg, Germany$^a$ \\
% ...
$ ^{14}$ II. Institut f\"ur Experimentalphysik, Universit\"at Hamburg,
                                     Hamburg, Germany$^ a$  \\
% ...
$ ^{18}$ Institute of Experimental Physics, Slovak Academy of
                Sciences, Ko\v{s}ice, Slovak Republic$^ f$\\
% ...
$ ^{25}$ Institute for Theoretical and Experimental Physics,
                                                 Moscow, Russia \\
 $ ^{26}$ Lebedev Physical Institute, Moscow, Russia$^ f$ \\
% ...
$ ^{28}$ LAL, Universit\'{e} de Paris-Sud, IN2P3-CNRS,
                            Orsay, France\\
 $ ^{29}$ LPNHE, Ecole Polytechnique, IN2P3-CNRS,
                             Palaiseau, France \\
% ...
$ ^{36}$ DESY, Institut f\"ur Hochenergiephysik,
                              Zeuthen, Germany$^ a$\\
 $ ^{37}$ Institut f\"ur Teilchenphysik,
          ETH, Z\"urich, Switzerland$^ i$\\
% ...
$ ^a$ Supported by the Bundesministerium f\"ur
        Forschung und Technologie, FRG,
        under contract numbers 6AC17P, 6AC47P, 6DO57I, 6HH17P, 6HH27I,
        6HD17I, 6HD27I, 6KI17P, 6MP17I, and 6WT87P \\
% ...
$ ^f$ Supported by the Deutsche Forschungsgemeinschaft\\
% ...
$ ^i$ Supported by the Swiss National Science Foundation\\
\vfill
\clearpage


\section{Introduction} \label{sect:intro}

\begin{figure}[htbp]
\vspace{-1.cm}
\begin{center}
\epsfig{file=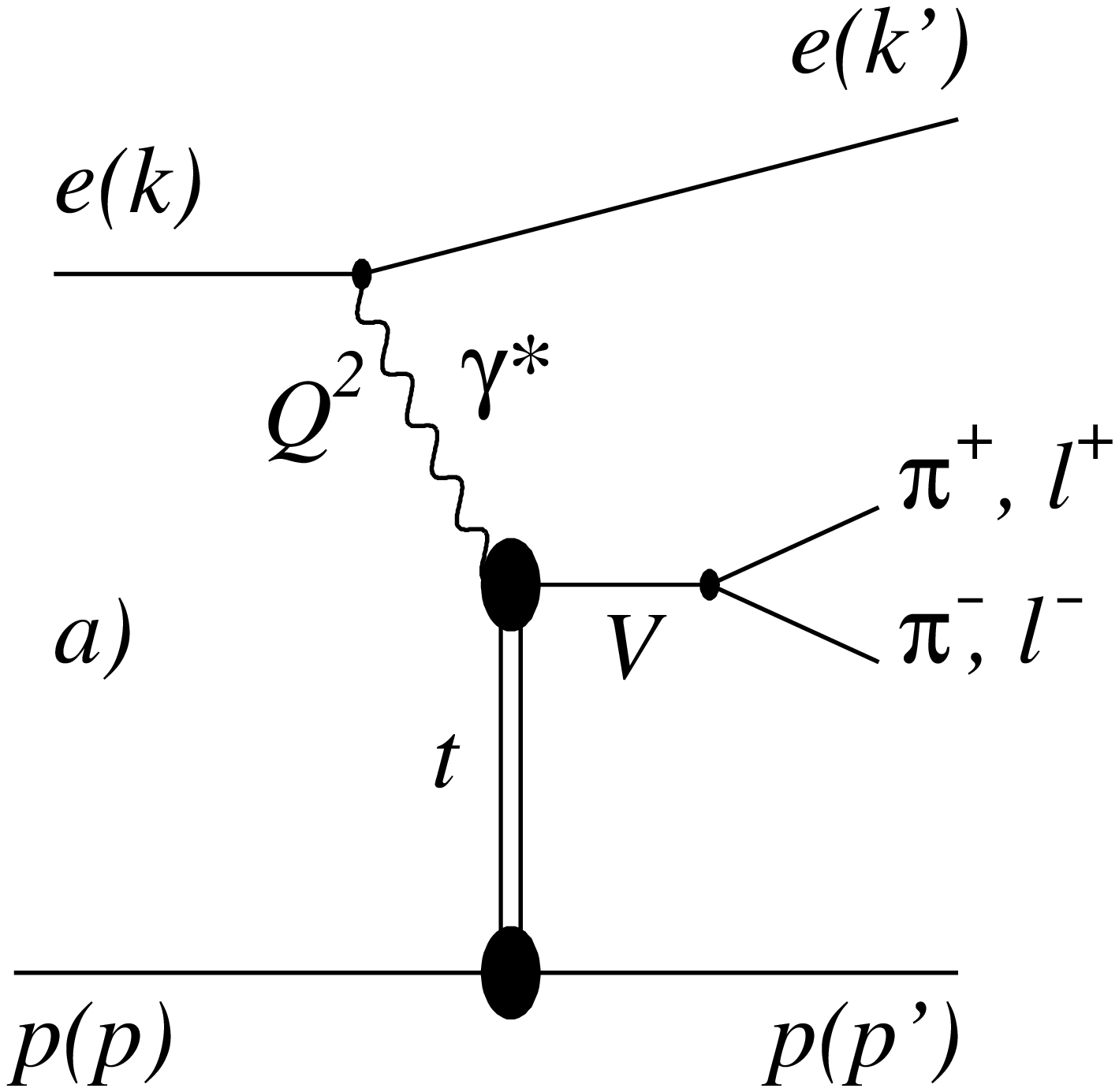,width=6cm,height=6cm}\epsfig{file=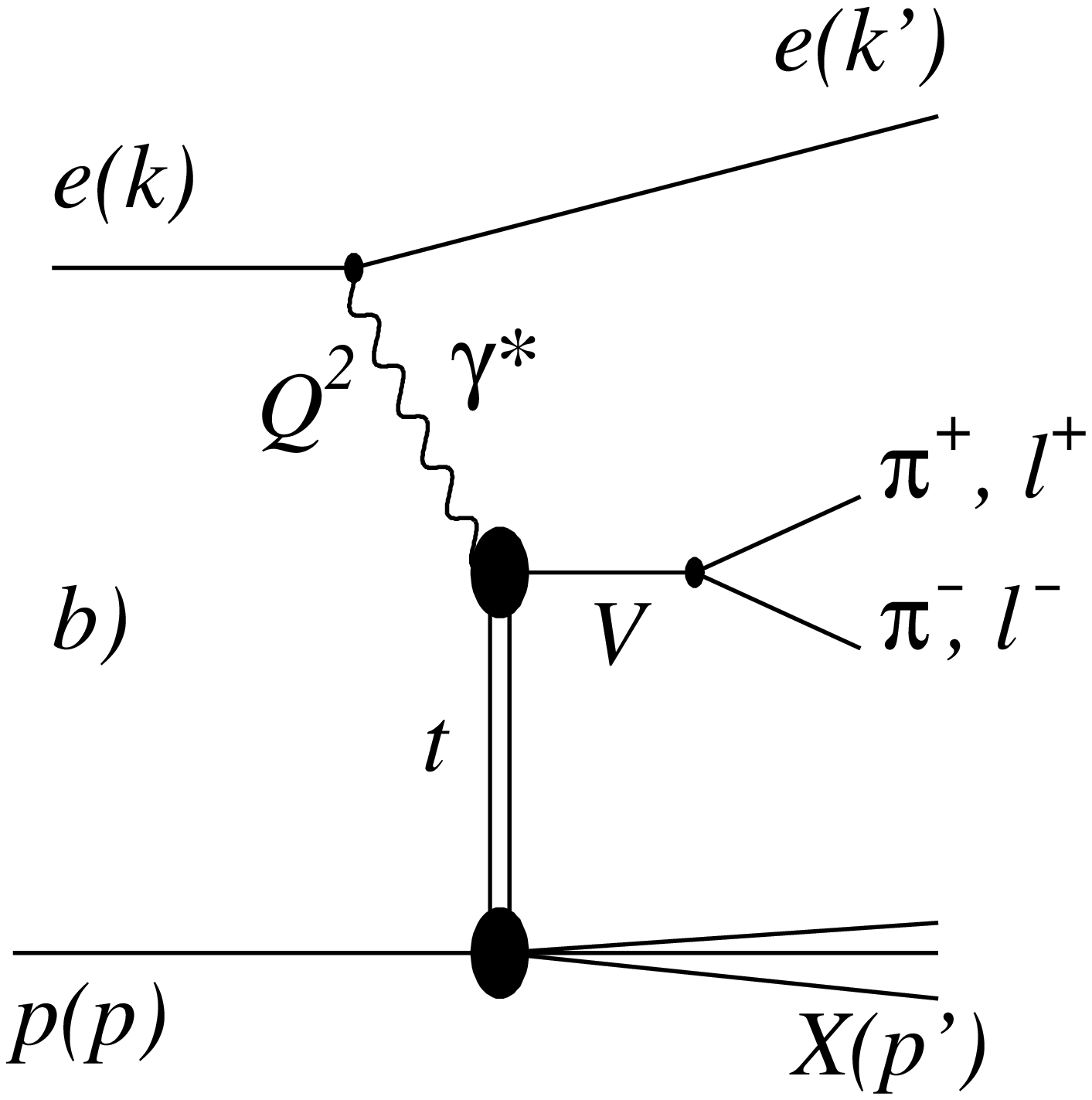,width=6cm,height=6cm}
\end{center}
\vspace{-1.cm}
\caption{Diffractive vector meson production: a) elastic; b) proton
dissociation.}
\label{fig:diag}
\end{figure}

The study of elastic production of vector mesons in photo- 
and leptoproduction (Fig. \ref{fig:diag}a) in fixed target experiments has 
provided information on the hadronic component of the photon  
and on the nature of diffraction. 

With the advent of the electron-proton collider HERA, there is renewed interest 
in vector meson production, in particular at large \qsq\ (\qsq\ is minus the square 
of the exchanged photon four-momentum).
HERA experiments have observed in deep-inelastic scattering that the proton 
structure function $F_2$ increases rapidly \cite{F2} with increasing \W, 
the \gsp\ invariant mass, which is in striking contrast with the slow rise of the 
total \gp\ cross section at $\qsq \simeq 0$ \cite{totalgammaxsection}.
In addition, the rise of $F_2$ already at relatively low \qsq\ values ($\qsq\ \lsim\ 2$ 
\gevsq) indicates that the transition between these two behaviours is rapid 
\cite{F2_low_Q2}.
The study of the elastic production of vector mesons is expected to provide useful 
information about these different regimes, and in particular about the transition 
between them.

The subject of this paper \cite{Bxl_papers} is the study of elastic \rh\ and 
\jpsi\ meson electroproduction at large \qsq\ ($\qsq > 8 $ \gevsq) 
and high \W\ (\av{\W} $\simeq 90$ \gevcsq), in the reactions 
\begin{equation}
e \ p \rightarrow e \ \rh \ p \\ ;\ 
\rh \rightarrow \pi^+ \ \pi^-,    \label{eq:rh}
\end{equation}
\begin{equation}
e \ p \rightarrow e \ \jpsi \ p \\ ;\ 
\jpsi \rightarrow l^+ \ l^- \\ ;\ 
l = e \ {\rm or} \ \mu.    \label{eq:jpsi}
\end{equation}

After the presentation of the event selection and of the mass distributions,
the total cross sections and the differential distributions for reactions 
(\ref{eq:rh}) and (\ref{eq:jpsi}) are studied.
Features relevant for the study of the transition regime are then discussed, 
in particular the energy dependence of the cross section, the \ttr\ distribution 
(\ttr\ is the square of the four-momentum transferred from the photon to 
the target), the vector meson polarisation, and the evolution of the 
\jpsi\ : \rh\ cross section ratio.
The results are discussed in the light of the interplay of hard and 
soft physics processes.
Results on \rh\ production in a similar kinematic range have been presented 
by the ZEUS Collaboration~\cite{ZEUS}.


\section{Models and phenomenology} \label{sect:models}

The elastic production of light vector mesons, in particular of \rh\ mesons,
by real or quasi-real photons ($\qsq \simeq 0$), a process called hereafter 
photoproduction, exhibits numerous features typical of soft, hadron-like
interactions.
These are a small angle peak in the distribution of the vector meson scattering 
angle with respect to the incident photon beam direction (``elastic peak''), 
a steepening of this distribution with increasing energy (``shrinkage''), 
for $W\ \gsim\ 10$ \gevcsq\ a slow increase with energy of the cross section, and 
s-channel helicity conservation ($SCHC$). 

These observations support the vector meson dominance model (VDM) \cite{VDM,Bauer}, 
according to which a photon with energy greater than a few \gev\ behaves predominantly 
as the superposition of the lightest (\rh, \om, \ph) $J^{PC}=1^{--}$ mesons.
In this framework, the cross section for elastic vector meson production is related 
to the total meson$-$proton cross section through the optical theorem, and the energy
dependence is related to the ``universal'' energy dependence of the total 
hadron$-$proton cross section.
Neglecting a contribution (``reggeon exchange'') which decreases with energy approximately 
as $s^{-0.5}$, the latter is parameterised at high energy as $\sigma_{tot} \propto 
\s^{\delta}$ with $\delta \simeq 0.0808$, $\sqrt{\s}$ being the hadron$-$proton 
centre of mass energy \cite {DL_hadronxsection}.
This is expressed in Regge theory as due to soft pomeron exchange.
The elastic production of \rh\ mesons has been studied extensively by fixed target 
experiments in photoproduction \cite{Bauer,gp_rho}, for intermediate \qsq\ \cite{moder_qsq_rho} 
and for \qsq\ $\gsim\ 6$ \gevsq\ \cite{EMC_NMC_rho}.
Also at HERA, the photoproduction of \rh\ mesons exhibits the characteristic features 
of soft interactions \cite{H1_Z_g_rho}.

In contrast, the VDM approach does not give a satisfactory description of \jpsi\ 
photoproduction data \cite{gp_jpsi,moder_qsq_jpsi,EMC_jpsi}: the cross section is smaller 
than the VDM prediction (see \cite{DL_1995}) and it increases significantly faster with energy 
than expected from soft pomeron exchange. 
This is also observed by the HERA experiments \cite{H1_Z_g_jpsi}.

Using the hard scale provided by the mass of the charm quark, the fast increase 
of the \jpsi\ photoproduction cross section was predicted in the framework of QCD
by Ryskin \cite{Ryskin}.
In his model, the interaction between the proton and the $c \bar c$ pair in 
the photon is mediated by a gluon ladder, and non-perturbative effects are 
included in the nucleon parton distribution. 

At high energy and high \qsq, \rh\ meson production is modelled using two different 
approaches based on QCD, which both refer to the pomeron as basically a two gluon system 
(see \cite{Low_Nussinov}). The \rh\ meson production is thus related to the
gluon distribution in the proton, but the two approaches emphasize respectively 
soft or hard behaviour.

In the soft model initially proposed by Donnachie and Landshoff 
\cite{DL_rho,Cudell,DL_1995}, the gluons are described 
non-perturbatively and the elastic cross section is expected to increase slowly with 
energy: $d\sigma / dt\ (t=0) \propto W^{4 \delta}$, with $\delta \simeq 0.0808$. 
The vector mesons are predicted to be mostly longitudinally polarised, the \gsp\
cross section to fall as $Q^{-6}$ and the \jpsi\ cross section to be comparable to 
that of the \rh.

In the second approach, perturbative QCD calculations similar to the work of Ryskin 
have been performed by several groups \cite{Kope, Brodsky, Ginzburg},
a hard scale being provided by the photon virtuality.
A major prediction of this model is a rapid increase of the cross section with \W, 
as a consequence of the rise of the gluon distribution in the proton. 
It is stressed, however, that non-perturbative processes are also expected to be
present in an intermediate energy region \cite {Kope,F_K_S}.
The scattering amplitude is obtained in these calculations from the convolution of 
the hadronic wave function of the photon, of the scattering amplitude of this hadronic 
component, and of the final state vector meson wave function.
This is because the photon is viewed at sufficient energy as the coherent superposition 
of hadronic states formed well before the target (essentially $q \bar q$ pairs), whereas
the final state meson is formed beyond it, the corresponding time scales 
being much longer than the interaction time. 
As a consequence, the spatial dimensions of the hadronic wave function are an essential 
parameter in the interplay of perturbative and non-perturbative effects.
It is predicted that the hard effects should show up earlier for small size 
objects than for large ones, and for longitudinally than for transversely 
polarised photons.
The \qsq\ dependence of the \gsp\ cross section is predicted, as in the non-perturbative 
approach, to fall as $Q^{-6}$ but when the evolution of the parton distribution  and
quark Fermi motion are taken into account, it was pointed out that the \qsq\ distribution 
is expected to be harder \cite{F_K_S}.
In view of the more compact \jpsi\ wave function,
the \jpsi\ : \rh\ cross section ratio has been predicted \cite{F_K_S} 
to exceed, at very high energy, the value 8\ : 9 obtained from SU(4) and the quark 
counting rule \cite{quark-counting_rule}.

In addition to these two ``microscopic'' approaches inspired by QCD, calculations are
also performed for elastic vector meson production 
on the basis of multiple pomeron exchange, with the effective pomeron intercept 
depending on the photon virtuality \cite{Kaidalov}. 
A stronger increase of the cross section with energy is again predicted than for 
soft pomeron exchange with $\delta \ \simeq 0.0808$.


\section{Detector and event selection} \label{sect:detector}

The data presented here correspond to an integrated luminosity of
2.8 \pbinv\ for the \rh\ and 3.1 \pbinv\ for the \jpsi\ mesons. 
They were collected in 1994 using the H1 detector. 
HERA was operated with 27.5 \gev\ positrons and 820 \gev\ 
protons\footnote {\ For the \rh\ studies only positron data are used, whereas
the small amount of data taken with $e^-p$ scattering is included in the \jpsi\ 
sample; in this report, {\it positrons} refers both to positrons and electrons.}. 
The detector is described in detail in ref.  \cite{detector}.

The event final state corresponding to reactions (\ref{eq:rh}) and (\ref{eq:jpsi}) 
consists of the scattered positron and two particles of opposite charges, originating 
from a vertex situated in the nominal $e^+p$ interaction region. 
In most cases, the scattered proton remains inside the beam pipe because of 
the small momentum transfer to the target in elastic interactions.

In the \qsq \ range studied here, the positron is identified as an 
electromagnetic cluster with an energy larger than 12 GeV, reconstructed in 
the backward electromagnetic calorimeter (BEMC) and associated with a hit 
in the proportional chamber (BPC) placed in front of it at 141 cm 
from the nominal interaction vertex\footnote {\ The forward ($+z$) direction,
with respect to which polar angles are measured, is defined as that of the incident 
proton beam, the backward direction is that of the positron beam.}.
The BEMC covers the polar angles $151^\circ <\rm \theta < 176^\circ$.
Its electromagnetic energy resolution is $\sigma_{E}/E \simeq 6 - 7$\%
in the energy range of the positrons selected for the present studies. 
The BPC angular acceptance is $155.5^\circ < \theta < 174.5^\circ$.
The scattered positron polar angle $\rm{\theta_e}$ is determined from the positions of 
the BPC hit and of the interaction vertex. 
The trigger used for the present analyses requires the presence of a total 
energy larger than 10 \gev\ deposited in the BEMC, outside a square of 
$32 \times 32\ {\rm cm^2}$ around the beam pipe. 
Additional cuts, similar to those used for the structure 
function analysis [1a], are applied to the hit and cluster 
position and shape in order to provide high trigger efficiency and good 
quality positron measurement. 
These cuts are complemented by the selection of events with 
\qsq \ $>$ 8 \gevsq. 

The decay pions (\rh\ events) or leptons (\jpsi\ events) are detected in the 
central tracking detector, consisting mainly of two coaxial cylindrical drift
chambers, 2.2 m long and respectively of 0.5 and 1 m outer radius. 
The charged particle momentum component transverse to the beam direction is 
measured in these chambers by the track curvature in the 1.15 T magnetic 
field generated by the superconducting solenoid which surrounds the 
inner detector, with the field lines directed along the beam axis. 
Two polygonal drift chambers with wires perpendicular to the beam direction, 
located respectively at the inner radius of the two coaxial chambers, 
are used for a precise measurement of the particle polar angle. 
For the present analyses, two tracks with transverse momenta $p_{t}$ larger 
than 0.1 GeV/c are required to be reconstructed in the central region of the 
tracker. 
For \rh\ production, the polar angles 
must lie in the range $25^\circ < \theta < 155^\circ$, 
corresponding to particles completely crossing the inner cylindrical 
drift chamber for interactions at the nominal vertex position. For \jpsi\ 
production, the accepted range is extended to $20^\circ < \theta < 160^\circ$
in order to increase the statistics as much as possible while keeping good
detection efficiency. 
The vertex position is reconstructed using these tracks.
No other track linked to the interaction vertex is allowed in the tracking detector, 
except possibly the positron track. 
To suppress beam-gas interactions, the vertex must be reconstructed within 30 cm of the 
nominal interaction point in $z$, which corresponds to 3 times the width of the 
vertex distribution. 
In addition, the accepted events are restricted to the range 
$40 < \W < 140$ \gevcsq\ for \rh\ mesons and $30 < \W < 150$ \gevcsq\ 
for \jpsi\ mesons.

The tracking detector is surrounded by a liquid argon calorimeter situated 
inside the solenoid and covering the polar angular range 
$4^\circ < \rm \theta <  153^\circ$ with full azimuthal coverage. 
In the case of elastic interactions, the calorimeters should register only activity
associated with the decay particles or the positron. 
However, due to noise in the calorimeters and to the small pile-up from different events, 
elastic \rh\ and \jpsi\ 
production can be accompanied by the presence of additional energy clusters.
These are considered in terms of the variable \eclmax, defined as 
the energy of the most energetic cluster which is not associated with a track. 
The \eclmax\ distribution shows a peak at small values, attributed mostly to elastic and 
diffractive interactions, and a broad maximum at higher energies.
The cut \eclmax\ $<$ 1 \gev\ is applied to enhance exclusive \rh\ production.
This is discussed in section \ref{sect:mass}, together with the effect of
the cut \modt $<$ 0.5 \gevsq.

The sample of events containing a positron and a vector meson candidate
includes two main contributions: 
elastic production (see Fig. \ref{fig:diag}a), defined by reactions 
(\ref{eq:rh}) and (\ref{eq:jpsi}), and events where the proton is diffractively excited 
into a system $X$ of mass $M_X$, which subsequently dissociates (Fig. \ref{fig:diag}b).
Non-resonant background is also present.
It is possible to identify most of the ``proton dissociation'' events with the components
of the H1 detector placed in the forward region \cite{diffrpaper}, namely the forward part of 
the liquid argon calorimeter $4^\circ \leq \theta \leq 10^\circ$, the forward muon detectors 
(arrays of muon chambers placed around the beam pipe in the proton direction,
$3^\circ \leq \theta \leq 17^\circ$) and the proton remnant tagger (an array of 
scintillators placed 24~m downstream of the interaction point,
$0.06^\circ \leq \theta \leq 0.17^\circ$). 
When particles from the diffractively excited system interact in the beam pipe 
and the collimators, the interaction products can be detected in these forward detectors.
The events are tagged as due to proton dissociation by the presence of a cluster with energy 
$E^{LAr}_{fw}$ larger than 1 \gev\ (0.75 \gev\ for the \jpsi\ candidates) at an angle 
$\theta^{LAr}_{fw} < 10 ^\circ$ in the liquid argon calorimeter, or by at least 2 pairs of hits 
in the forward muon detectors (one hit pair is compatible with noise),
or by at least one hit in the proton remnant tagger. 
This last criterion is not used for the \jpsi\ events, since they have a flatter 
\ttr\ distribution than the \rh\ events and would be partially vetoed by the proton tagger.

In order to minimise the effects of QED radiation in the initial state, the difference 
between the total energy and the total longitudinal momentum \eminpz\ of the positron 
and the two particles emitted in the central part of the detector is required 
to be larger than 45 GeV. If no particle, in particular a radiated photon, has 
escaped detection in the backward direction, \eminpz\ should be twice the
incident positron energy, i.e. 55 \gev.

The selection criteria for the two samples, supplemented by the mass
selections discussed in section \ref{sect:mass}, are summarised in Table \ref{tab:sel}.

\begin{table}[htbp]
\begin{center}
\begin{tabular}{|l|c|c|}
\hline
   \multicolumn{1}{|c|}{} & 
     \multicolumn{1}{|c|}{\rh} & \multicolumn{1}{|c|}{\jpsi} \\
\hline
\hline
   \multicolumn{1}{|l|}{positron selection} & 
            \multicolumn{2}{|c|}{em. cluster $>$ 10 GeV in BEMC outside $32 \times 32\ {\rm cm^2}$} \\
   \multicolumn{1}{|c|}{} & 
      \multicolumn{2}{|c|}{reconstructed positron energy $>$ 12 GeV} \\
    \multicolumn{1}{|c|}{} & 
 	 \multicolumn{2}{|c|}{associated BPC hit} \\
     \multicolumn{1}{|c|}{} & 
         \multicolumn{2}{|c|}{\qsq\ $>$ 8 \gevsq} \\
\hline
   \multicolumn{1}{|l|}{VM reconstruction} & 
         \multicolumn{2}{|c|}{2 tracks fitting to vertex (+ possibly $e^+$)} \\
     \multicolumn{1}{|c|}{} & 
         \multicolumn{2}{|c|}{$p_{t} > 0.1$ \gevc }   \\   
     \multicolumn{1}{|c|}{} & 
         \multicolumn{1}{|l|}{$25^\circ<\theta_{track}<155^\circ$}  & 
           \multicolumn{1}{|l|}{$20^\circ<\theta_{track}<160^\circ$} \\
     \multicolumn{1}{|c|}{} & 
         \multicolumn{1}{|l|}{$40 < \W < 140$ \gevcsq } & 
             \multicolumn{1}{|l|}{$30 < \W < 150$ \gevcsq} \\

\hline
   \multicolumn{1}{|l|}{background suppression} & 
         \multicolumn{2}{|c|}{$|z_{vertex} - z_{nom.}| < 30$ cm} \\
     \multicolumn{1}{|c|}{} & 
         \multicolumn{1}{|l|}{\eclmax\ $<$ 1 \gev} &  \multicolumn{1}{|c|}{$-$}  \\
     \multicolumn{1}{|c|}{} & 
         \multicolumn{1}{|l|}{\modt\ $<$ 0.5 \GeVsq}    & \multicolumn{1}{|c|}{$-$} \\
     \multicolumn{1}{|l|}{p. dissoc. ev. tagging} & 
         \multicolumn{1}{|l|}{$E^{LAr}_{fw} < 1\ \gev\ {\rm for}\ \theta^{LAr}_{fw} < 10^\circ$}  & 
             \multicolumn{1}{|l|}{$E^{LAr}_{fw} < 0.75\ \gev\ {\rm for}\ \theta^{LAr}_{fw} < 10^\circ$} \\
    \multicolumn{1}{|c|}{} & 
         \multicolumn{2}{|c|}{$\leq 1$ hit pair in forward muon detectors} \\
     \multicolumn{1}{|c|}{} & 
         \multicolumn{1}{|l|}{no hit in proton tagger}  &     \multicolumn{1}{|c|}{$-$} \\
\hline
   \multicolumn{1}{|l|}{radiative corrections} & 
         \multicolumn{2}{|c|}{\eminpz\ $>$ 45 \gev } \\
\hline
   \multicolumn{1}{|l|}{mass selection} & 
         \multicolumn{1}{|l|}{0.6 $<$ \mpipi\ $<$ 1.0 \gevcsq} & 
             \multicolumn{1}{|l|}{2.8 $<$ \mll\ $<$ 3.4 \gevcsq} \\\hline
\hline
  \multicolumn{1}{|l|}{selected sample} & 
         \multicolumn{1}{|c|}{180 events}  & 
             \multicolumn{1}{|c|}{31 events} \\
\hline
\end{tabular}
\end{center}
\caption{Selection criteria for \rh\ and \jpsi\ events.}
\label{tab:sel}
\end{table}


\section{Kinematics and cross section definitions} \label{sect:kin}

The kinematics of reactions (\ref{eq:rh}) and (\ref{eq:jpsi}) are described with the 
variables commonly used for deep-inelastic interactions. In addition to \s\ (the square 
of the $e^+p$ centre of mass energy), \qsq\ and \W, it is useful to define the 
two Bjorken variables $ y= p \cdot q / p \cdot k $ (in the proton rest frame, the energy fraction 
transferred from the positron to the hadrons) and $ x= \qsq  / 2p \cdot q $, where 
$k$, $p$, $q$ are, respectively, the four-momenta of the incident positron, 
of the incident proton and of the virtual photon. 
These variables\footnote{ 
\ In this paper, the positron and proton masses are neglected.} 
obey the relations $\qsq = x y s $ and $W^2 = \qsq\ (\frac{1}{\x} -1)$. 

The kinematical variables can be reconstructed from four measured quantities: 
the energies and the polar angles of the scattered positron and of the vector meson. 
With the ``double angle'' method \cite{Koijman_Workshop} used for the present analyses, 
\qsq\ and \y\ are computed using the polar angles $\theta$ and $\gamma$ of the positron 
and of the vector meson, which are well measured:
\begin{equation}
\qsq = 4 E_0^2 \ \frac {\sin\gamma \ (1 + \cos \theta)}
       {\sin\gamma + \sin\theta - \sin(\gamma + \theta)} ,    
                                                 \label{eq:qsq}
\end{equation}
\begin{equation}
y = \frac {\sin\theta \ (1 - \cos \gamma)}
       {\sin\gamma + \sin\theta - \sin(\gamma + \theta)} ,
                                                \label{eq:y}
\end{equation}
where $E_0$ is the energy of the incident positron.

The meson momentum components are obtained from the measured decay products.
The momentum of the scattered positron is computed from \qsq\ and \y, which provides
better precision than the direct measurement. 
The energy transfer to the proton being negligible, the absolute value of \ttr\ 
is given by:
\begin{equation}
\modt \simeq (\vec{p}_{tp})^{2} = (\vec{p}_{te} + \vec{p}_{tv})^2,    \label{eq:t}
\end{equation}
where $\vec{p}_{tp}$, $\vec{p}_{te}$ and $\vec{p}_{tv}$ are, respectively, 
the momentum components transverse to the beam direction of the final state proton, 
positron and vector meson\footnote{\ The lowest \modt\ value kinematically allowed, 
$t_{min} \simeq (\qsq + m^2_{V})^2 \ m^2_p\ / y^2 s^2$, is negligible in this 
experiment.}.

The fourth quantity which is directly measured, the positron energy, is used to compute 
the variable \eminpz:
\begin{equation}
E-p_z = (E_e + E_v) - (p_{ze} + p_{zv}),    \label{eq:eminpz}
\end{equation}
$E_e$ and $E_v$ being the energies of the scattered positron and of the 
vector meson, and $p_{ze}$ and $p_{zv}$ their momentum components 
parallel to the beam direction.


The cross section for elastic electroproduction of a vector meson $V$
can be converted into a  \gsp\ cross section using the relation
\begin{equation}
\frac{d^2 \sigma_{tot} (ep \rightarrow e V p)}{dy \ dQ^2} =
\Gamma \ \sigma_{tot} (\gamma ^*p \rightarrow V p) = \
\Gamma \ \sigma_{T} (\gamma ^*p \rightarrow V p) \ (1 + \varepsilon \ \R),   
                                            \label{eq:sigma}
\end{equation}
where $\sigma_{tot}$, $\sigma_{T}$ and $\sigma_{L}$ are the total, transverse 
and longitudinal \gsp\ cross sections, 
\begin{equation}
R\ = \sigma_{L} / \sigma_{T},
                                             \label{eq:Rdef}
\end{equation}
and $\Gamma$ is the flux of transverse virtual photons given by
\begin{equation}
\Gamma = \frac {\alpha_{em} \ (1-\y+\y^2/2)} {\pi\  \y\  \Qsq};
                                             \label{eq:flux}
\end{equation}
$\varepsilon$\ is the polarisation parameter
\begin{equation}
\varepsilon = \frac{1 - \y}{1-\y+\y^2/2}.
                                       \label{eq:epsil}
\end{equation}
%


Information on the vector meson production process can be obtained from
the angular distributions of the decay particles. 
In particular, the probability \rzzzz\ for the \rh\ meson to be longitudinally 
polarised can be determined from the distribution of \costhst, where $\theta^*$ 
is the angle, in the \rh\ rest frame, between the direction of the positively 
charged decay pion and the \rh\ direction in the \gsp\ centre of mass system
(helicity frame) \cite{Bauer,Schilling_Wolf}:
\begin{equation}
\frac {{\rm d}N} {{\rm d}\costhst} \propto  1 - \rzzzz + (3 \ \rzzzz -1) \ \cos^2\theta^* .                               
                                           \label{eq:costhst}
\end{equation}
With the assumption of s-channel helicity conservation ($SCHC$), 
\rzzzz\ is related to \R:
\begin{equation}
\R = \frac{1}{\varepsilon} \ \frac{\rzzzz}{1-\rzzzz} .  
                                              \label{eq:R}
\end{equation}


\section{Mass distributions and final samples} \label{sect:mass}

\begin{figure}[t] \unitlength 1mm
\begin{picture}(160,80)
\put(1,0) {\epsfig{file=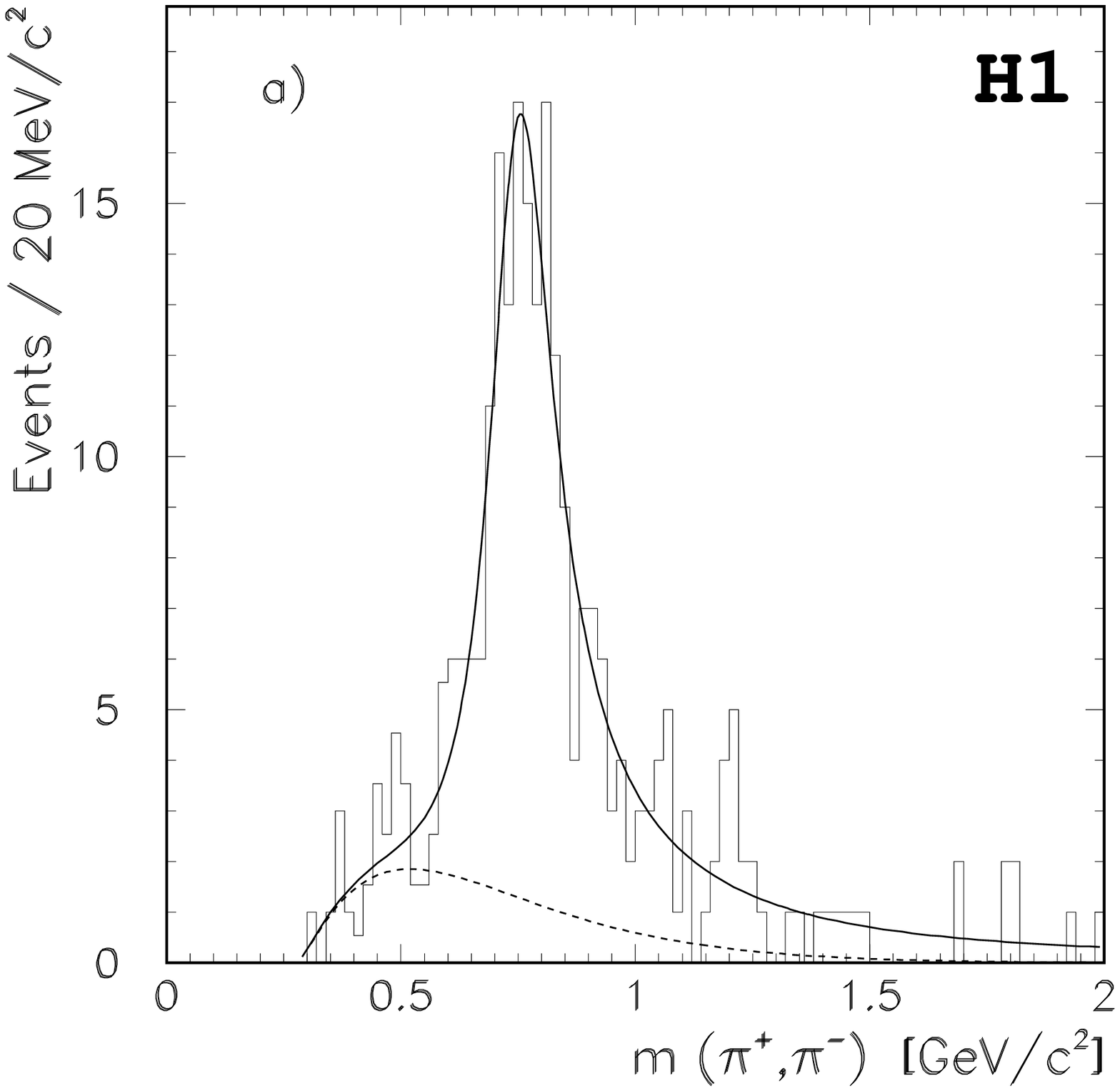,height=8cm,width=8cm}}
\put(39.63,36) {\epsfig{file=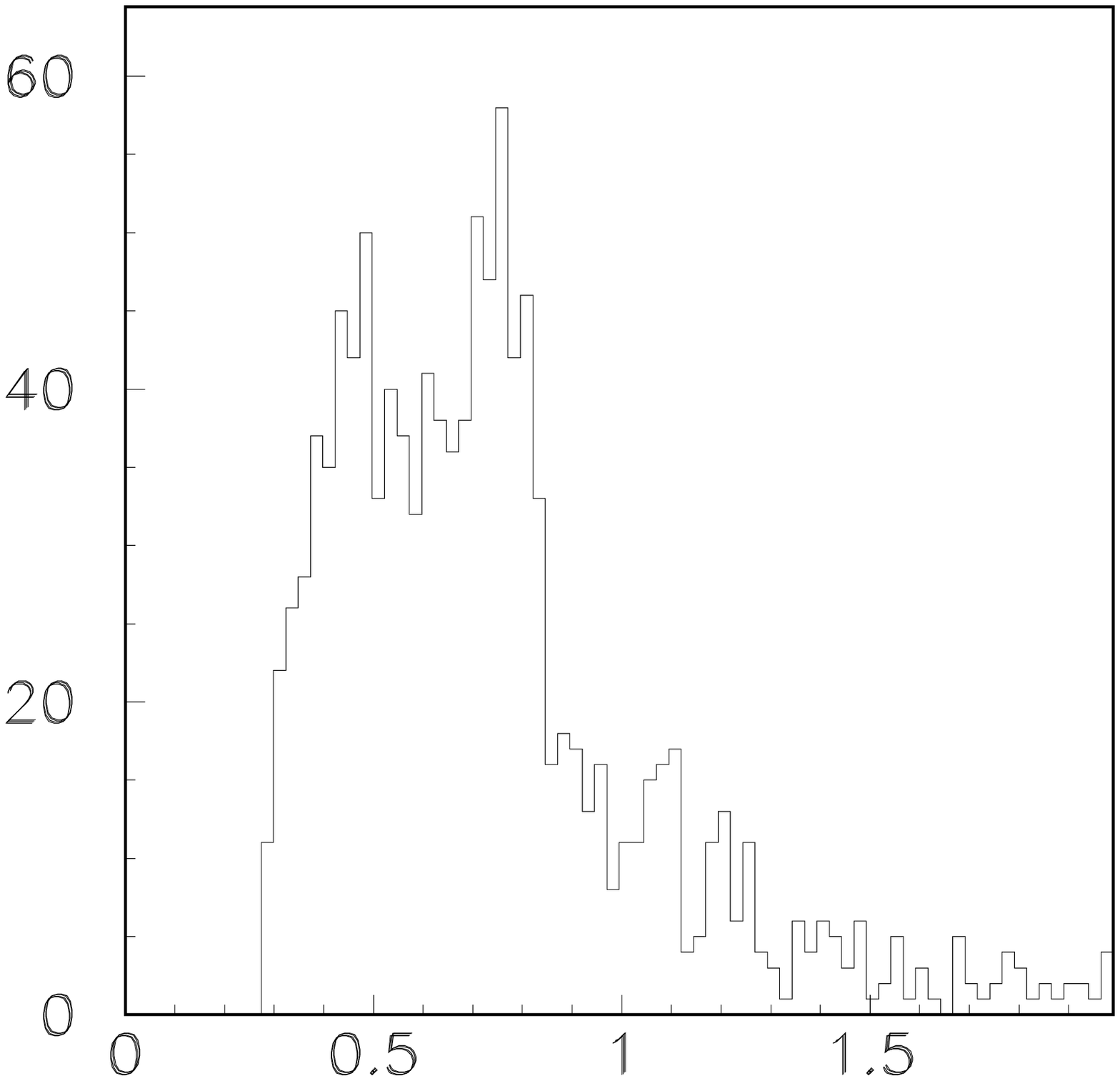,height=4cm,width=3.7cm}}
%
\put(80,0) {\epsfig{file=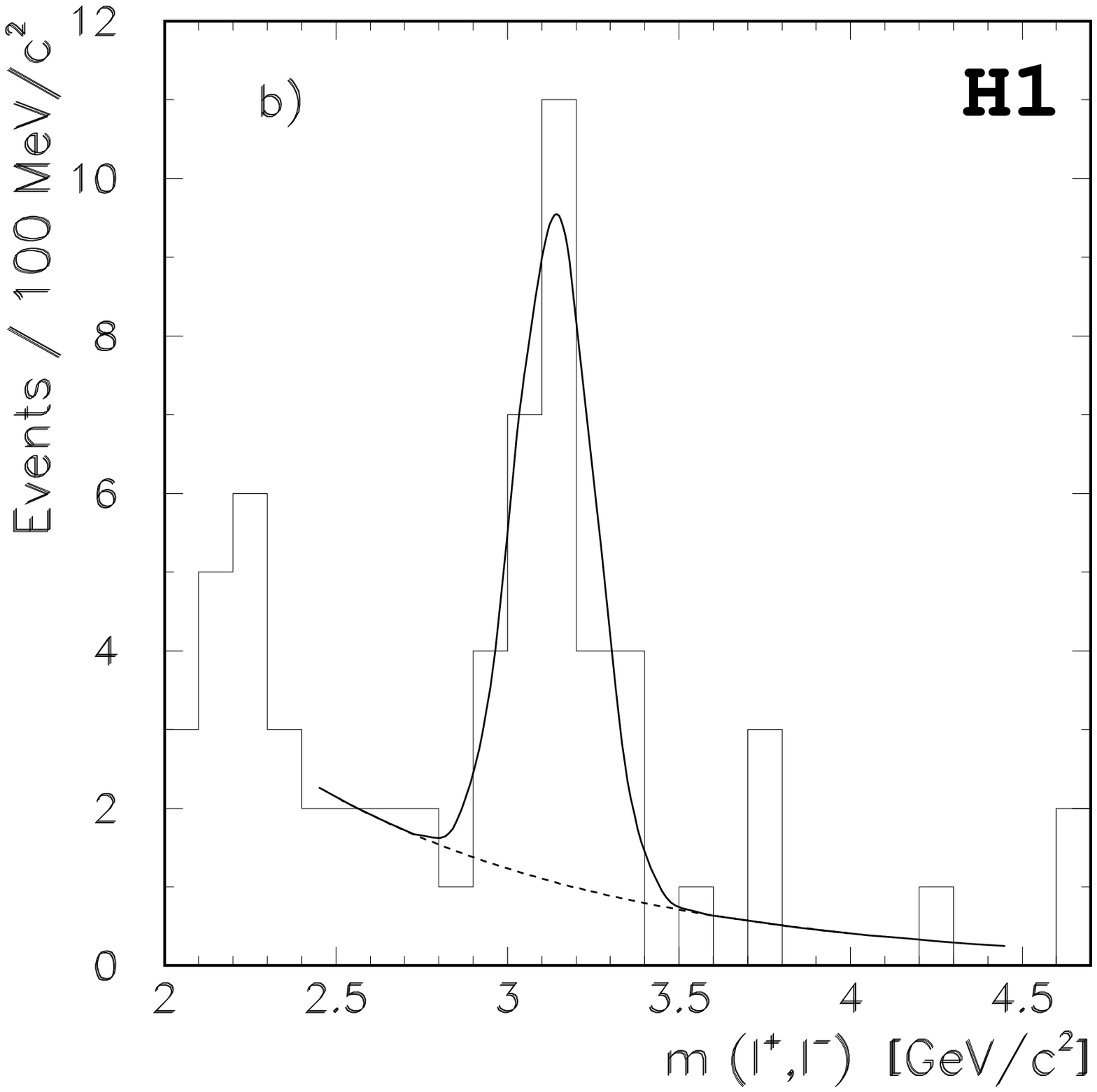,height=8cm,width=8cm}}

\end{picture}
%
\caption{a) \mpipi\ mass distribution for the events with \qsq\ $>$ 8 \gevsq,
$40 < \W < 140$ \gevcsq, \modt\ $<$ 0.5 \gevsq, \eclmax \ $<$ 1 \gev\ and 
no proton dissociation signal in the forward detectors; the superimposed 
curve is the result of a fit to a relativistic Breit-Wigner distribution over 
the background (\ref{eq:ps_bg}), which is described by the dashed curve. 
In the insert: same distribution without the \eclmax\ and \ttr\ cuts;
b) \mll\ mass distribution for the events with \qsq\ $>$ 8 \gevsq,
$30 < \W < 150$ \gevcsq\ and no proton dissociation signal in the forward 
detectors; the superimposed curve is the result of a fit to a Gaussian 
distribution over an exponential background in the range 2.4 $<$ \mll\ $<$ 4.5 
\gevcsq.}
\label{fig:mass}
\end{figure}

Fig. \ref{fig:mass}a shows for the selected events (Table \ref{tab:sel}) the 
distribution of the invariant mass \mpipi\ in the range \mpipi\ $<$ 2 \gevcsq, 
obtained by assigning the pion mass to the particles detected in the central 
tracker. 
The mass distribution without the \eclmax\ and \ttr\ cuts 
(insert in Fig. \ref{fig:mass}a) is seen to peak at small \mpipi\ values. 
The \eclmax\ cut strongly reduces the background of events containing
neutral particles, and enhances the \rh\ peak.
The \ttr\ cut is very effective in rejecting non-resonant events containing, in
addition to the \rh\ candidate and the positron, particle(s) with a significant 
transverse momentum, which is not used to compute $\vec{p}_{tp}$ in 
eq. (\ref{eq:t}). 
This cut also enhances the elastic production signal compared to the background 
of proton dissociation events, which are known to have a flatter \ttr\ distribution. 
In total, 180 events are found in the \rh\ peak region with $0.6 < \mpipi < 1.0$ 
\gevcsq.

%

In Fig. \ref{fig:mass}a, events compatible with the \ph\ mass, when the charged 
particles detected in the central tracker are considered as kaons, have been removed
($m_{K^+K^-} < 1.04$ \gevcsq). 
Assuming that vector mesons are produced according to the quark counting rule
with the SU(3) ratios
\rh\ : \om\ : \ph\ = 9\ : 1\ : 2, the Monte Carlo simulation described below
indicates that the remaining \ph\ and \om\ reflections contribute, in the \mpipi\ 
range (0.4 $-$ 0.6) \gevcsq, 2.0\% of the \rh\ signal in the peak, and 0.7\% 
in the range (0.6 $-$ 1.0) \gevcsq. 
These contributions were subtracted statistically.

The \mpipi\ distribution is described by a relativistic Breit-Wigner function, 
over a non-resonant background attributed to incompletely
reconstructed diffractive photon dissociation. 
The Breit-Wigner function has the form \cite{Jackson}
\begin{equation}
\frac {dN(\mpp)} {d\mpp} = \frac {\mpp \ \mrho \ \Gmpp}
               {(\mrhosq - \mppsq)^2 + \mrhosq \ \Gmppsq},
                                       \label{eq:b_w}
\end{equation}
with \Gmpp\ the mass dependent width
\begin{equation}
\Gmpp = \Grho \ (\frac {q^*} {q_0^*})^3 \ 
        \frac {2} {1 + (q^* / q^*_0)^2}.
                                       \label{eq:GJ}
\end{equation}
Here \mrho\ is the \rh\ resonance mass and \Grho\ the width; $q^*$ is the pion momentum 
in the ($\pi^+\pi^-$) rest frame, and $q^*_0$ this momentum when $\mpipi = \mrho$.

The background has been parameterised using the distribution  
\begin{equation}
\frac {dN(\mpp)} {d\mpp} = {\alpha_1 \ (\mpp - 2 \mpi)^{\alpha_2} \ e^{-\alpha_3 \mpp}},
                                       \label{eq:ps_bg}
\end{equation}
where \mpi\ is the pion mass and $\alpha_1$, $\alpha_2$ and $\alpha_3$ are free parameters.
This form, which includes a two pion threshold and an exponential fall off, is in 
qualitative agreement with the background shape in the insert of Fig. \ref{fig:mass}a.

With these parameterisations, the resonance mass is 763 $\pm$ 10 \mevcsq\ and
the width is $176 \pm 23$ \mevcsq, in agreement with the Particle Data Group (PDG) values 
of 770 and 151 \mevcsq\ \cite{PDG}.
No skewing to low values of \mpipi\ is needed to describe the \rh\ shape (for
the Ross-Stodolsky parameterisation \cite{RS}, the skewing exponent is found 
to be $n = 0.3 \pm 0.5$).

Two alternative forms have been used for the resonance width:
\begin{equation}
\Gmpp = \Grho \ (\frac {q^*} {q_0^*})^3 \ \frac {\mrho} {\mpp}
                                       \label{eq:Gmm}
\end{equation}

and
\begin{equation}
\Gmpp = \Grho \ (\frac {q^*} {q_0^*})^3.
                                       \label{eq:G1}
\end{equation}
They also give a \rh\ meson mass and width compatible with the PDG data.
%

The non-resonant background under the \rh\ peak is estimated to be 11 $\pm$ 6\%. 
The error includes the uncertainty on the resonance parameterisation and on 
the background shape, estimated by using for the latter an alternative linearly 
decreasing form.

The distribution of the invariant mass \mll\ for the selected events
in the \jpsi\ region is presented in Fig. \ref{fig:mass}b. 
The \jpsi\ mass is $3.13 \pm 0.03$ \gevcsq. The peak width is slightly larger than, but 
compatible with, the expectation obtained from the detector simulation. 
No \ttr\ selection is applied since the \ttr\ distribution is significantly flatter 
than for the \rh\ mesons (see Fig. \ref{fig:modt_1}). 
No \eclmax\ cut is required in view of the small background in this
high mass region (compare Fig. \ref{fig:mass}b and insert in Fig. \ref{fig:mass}a): 
all \jpsi\ candidate events have \eclmax\ $<$ 1.35 \gev, of which 4 have \eclmax\
larger than 1 \gev.

A sample of 31 \jpsi\ candidate events is thus selected with the cut 
$| \mll -m_{\psi} |$ $< 300\ {\rm MeV/c^2}$, where $m_{\psi}$ is the \jpsi\ meson mass. 
The non-resonant background is estimated by fitting the sidebands using an 
exponential distribution and amounts to roughly 20\% (6.8 events). 
No lepton identification is required, but 10 of the \jpsi\ candidate events contain two  
identified electrons and 7 contain two identified muons.

One event with \eclmax\ $>$ 1 \gev\ is a $\psi^\prime \rightarrow \jpsi \pi^0 \pi^0$ 
candidate, with one identified muon and neutral clusters detected in the 
electromagnetic part of the liquid argon calorimeter, attributed to the interaction of 
photons from $\pi^0$ meson decay.
The measured \jpsi\ mass is $3.16 \pm 0.04$ \gevcsq. 
The invariant mass computed using the two charged tracks and the neutral 
clusters is $3.67 \pm 0.09$ \gevcsq, in excellent agreement with the Particle Data Group 
value (3.69 \gevcsq) \cite{PDG}.

Kinematical characteristics of the selected events are summarised in Table
\ref{tab:events}.

\begin{table}[htbp]
\begin{center}
\begin{tabular}{|l|c|c|}
\hline
\                                                & \rh                & \jpsi              \\
\hline
\hline
\ \av {\qsq} [\gevsq]                            & $13.4 \pm 0.4$     & $17.7 \pm 1.5$     \\
\ \av {\W} [\gevcsq]                             & $81 \pm 2$         & $92  \pm 6$        \\
\ \av {{\rm scattered}\ e^+\ {\rm energy}}[\gev] & $25.5 \pm 0.1$     & $24.9 \pm 0.3$     \\
\ \av {{\rm scattered}\ e^+\ p_t} [\gevc]        & $3.4 \pm 0.1$      & $3.9 \pm 0.2$      \\
\ \av {{\rm meson\ energy}} [\gev]               & $4.2 \pm 0.1$      & $6.0 \pm 0.3$      \\
\ \av {{\rm track}\ p_t} [\gevc]                 & $1.7 \pm 0.1$      & $2.3 \pm 0.2$      \\
\hline               
\end{tabular}
\end{center}
\caption{Averages of kinematical variables characterising the selected events.}
\label{tab:events}
\end{table}

\section{Corrections and simulations} \label{sect:sim}

Table \ref{tab:corr} summarises the correction factors applied to the selected 
samples to take account of detector acceptance and efficiencies, smearing effects, 
losses due to the selection criteria and remaining backgrounds. 

Most corrections are estimated using a Monte Carlo simulation based on the 
vector meson dominance model, which permits variation of the \qsq, 
\W\ and \ttr\ dependences, as well as of the value of \R\ \cite{Benno}. 
The H1 detector response is simulated in detail, and the events are subjected
to the same reconstruction and analysis chain as the data. 

The accuracies of the \rh\ (\jpsi) variable measurements are for \W, 3.3 (4.3) \gevcsq, 
for \qsq, 0.4 (0.4) \gevsq, for \ttr, 0.06 (0.10) \gevsq.
The values of the \ttr\ slopes are little affected by the detector resolution.

The scattered positron selection criteria induce \qsq\ dependent losses for 
\qsq\ $\leq$ 12 \gevsq.
The error quoted in Table \ref{tab:corr} corresponds to a systematic uncertainty on the 
positron direction of 2 mrad. 
The charged track selection criteria induce \W\ dependent losses for small and high \W,
depending on the accepted \W\ range in the two selections.
A $\qsq-\W$  correlation of the losses is observed, and taken into account
in the corrections.
The correction for the \ttr\ cut in the \rh\ sample is computed using the 
measured \ttr\ slope (see section \ref{sect:t}).

\begin{table}[htbp]
\begin{center}
\begin{tabular}{|l|c|c|}
\hline
\                                     & \rh                & \jpsi           \\
\hline
\hline
\ trigger                             & \multicolumn{2}{|c|} {$1.01 \pm 0.02$} \\
\ positron acceptance (\qsq\ dep.)    & $1.16 \pm 0.03$    & $1.15 \pm 0.03$ \\
\ BPC hit $-$ cluster link              & \multicolumn{2}{|c|} {$1.03 \pm 0.02$} \\
\ tracker acceptance (\W\ dep.)       & $1.07 \pm 0.01$    & $1.29 \pm 0.03$ \\
\ track reconstr. (per track)         & $1.05 \pm 0.03$    & $1.03 \pm 0.03$ \\
\ track $p_{tmin}$ (per track)        & $1.02 \pm 0.01$    & $1.00 \pm 0.01$          \\
\ \modt\ cut                           & $1.03 \pm 0.02$    & \multicolumn{1}{|c|}{$-$} \\
\ \eminpz\ cut                         & $1.02 \pm 0.01$    & $1.01 \pm 0.01$          \\
\ \eclmax\ cut                         & $1.03 \pm 0.03$    & \multicolumn{1}{|c|}{$-$} \\
\ forward det. cuts (\modt\ dep.)     & $1.04 \pm 0.02$    & $1.03 \pm 0.03$ \\
\ mass selection                      & $1.22 \pm 0.01$    & $1.00 \pm 0.02$ \\
\hline
\hline
\ non-resonant background             & $0.89 \pm 0.06$    & $0.78 \pm 0.14$ \\
\ proton dissoc. background           & $0.91 \pm 0.08$    & $0.75 \pm 0.11$ \\
\ $\phi$ and $\omega$ background      & $0.99 \pm 0.01$    & \multicolumn{1}{|c|}{$-$} \\
\hline
\hline
\ photon flux / bin integration       & $1.00 \pm 0.04$    & $1.00 \pm 0.07$ \\
\ radiative corrections               & $0.96 \pm 0.03$    & $1.00 \pm 0.04$ \\
\ luminosity                          & \multicolumn{2}{|c|} {$1.00 \pm 0.02$}  \\
\hline
\end{tabular}
\end{center}
\caption{Correction factors and systematic errors, 
averaged over the data samples.}
\label{tab:corr}
\end{table}

The choice of the \eclmax\ cut for the \rh\ sample is a compromise between the 
loss of elastic events to which a cluster is accidentally associated in the calorimeter, 
and the presence of non-resonant background in the final sample. 
The loss is estimated using a Monte Carlo simulation which includes random 
noise in the calorimeters superimposed on elastic events. 

A simulation indicates that 2\% of the elastic \rh\ events with \modt\ $<$ 0.5 \gevsq\ 
are lost because the proton has acquired sufficient $\vec{p}_t$ to interact in the 
beam pipe walls, giving interaction products which are registered in the proton tagger; 
this loss is thus \ttr\ dependent. 
For the \jpsi\ sample, 1.5\% of the events are lost because of interaction products
registered in the forward muon detectors.
Another 2\% loss for both samples is due to spurious hits in the latter.

In view of the uncertainty on the high mass shape of the resonance, the \rh\ cross 
section is quoted in this paper for \mpipi\ $<$ 1.5 \gevcsq, 
i.e. $\simeq m_{\rho} + 5 \ \Gamma_{\rho}$. 
The uncertainty is larger in the present case than for low energy data, for which a 
natural cut off is imposed by the limited available energy.
With this definition of \rh\ resonance production line shape, the correction  for the mass 
selection 0.6 $<$ \mpipi\ $<$~1.0 \gevcsq\ is respectively
$21\%$ and $23\%$ for parameterisations (\ref{eq:GJ}) and (\ref{eq:Gmm}).

A simulation was performed in order to estimate the contribution to the final samples of 
proton dissociation events which are not tagged by the forward detectors. 
The distribution of the target mass $M_X$ is parameterised as $1/M_X^2$. 
High mass states are assumed to decay according to the Lund string model \cite{jetset} or, 
alternatively, to a final state with particle multiplicity following the KNO scaling law 
and isotropic phase space distribution.
In the resonance domain, the mass distribution follows measurements from $p$ dissociation 
on deuterium \cite{Goulianos} and resonance decays are described according to their 
known branching ratios. 
The decay particles are followed through the beam pipe walls and the forward detectors. 

For the \rh\ sample, the correction factor for the contamination of undetected proton 
dissociation events in the selected \rh\ sample is $0.91 \pm 0.08$. 
This number is obtained from the number of measured events tagged and not tagged by the forward 
detectors, and from the detection probabilities provided by the Monte Carlo simulation.
No assumption needs to be made for the ratio of proton dissociation to elastic events.
The error is a conservative estimate taking into account the uncertainties on the 
efficiencies of the forward detectors for tagging proton dissociation events and on the 
dissociation model. 

The correction factor for unobserved proton dissociation background in the selected
\jpsi\ sample, for which the proton tagger is not used, is $0.75 \pm 0.11$.

The cross section measurements are  given in the QED Born approximation for  
electron interactions. 
The effects of higher order processes are estimated using the HERACLES 4.4 generator 
\cite{Heracles}.

Radiative corrections for the \rh\ sample are of the order of 4\% after the cut 
\eminpz\ $>$ 45 \gev, and are weakly dependent on \qsq\ and \W. 
A systematic error of 3\% is obtained by varying the effective \qsq\ dependence 
of the \gsp\ cross section from $Q^{-4}$ to $Q^{-6}$ and by modifying the 
\W\ dependence from a constant to a linearly increasing form\footnote{\ In practice, 
the input to the program is an effective ``$F_2$ structure function'' 
parameterisation, with the chosen \qsq\ and \W\ dependences of the $ep$ cross section
for vector meson production.}.
The small value of the correction is due to the high \eminpz\ cut resulting from 
the good BEMC resolution; for the  chosen value of the cut, small smearing 
effects are observed.

For the \jpsi\ sample, the radiative corrections determined using the measured
\qsq\ and \W\ dependences of the cross section vary from $+2$\% to $-2$\%.


\section{Results} \label{sect:results}

\subsection{Electroproduction cross sections}
\label{sect:xsect}

The \rh\ and \jpsi\ data are grouped in several (\qsq, \W) bins. 
Table \ref{tab:xsect} gives, for each bin, the number of events, the integrated 
$ep$ cross section and the \gsp\ cross section obtained using relation 
(\ref{eq:sigma}) for a given ($Q_0^2,W_0$) value, taking into account the observed
dependence across the bin. 
All known smearing, acceptance and background effects are corrected for.
For the \rh\ sample, each event is weighted using the differential flux factor 
given by eq. (\ref{eq:flux}).
A 4\% systematic error accounts for the uncertainty in the \qsq\ 
and \W\ dependences of the cross section used for the bin size integration and the 
bin centre correction. 
For the \jpsi\ sample, in view of the small statistics, the photon flux is integrated 
over each (\qsq,\W) bin. Since the data span a large range in \qsq\ and \W, this 
leads to a systematic error on the cross section of the order of 7\%.

The integrated cross section for \rh\ meson electroproduction 
with \mpipi\ $<$ 1.5 \gevcsq\ is 
\begin{equation}
\sigma(e\ p \rightarrow e\ \rh\ p) = 96 \pm 7\ (stat.) \pm 13\ (syst.)\ {\rm pb},
   \label{eq:xrho}
\end{equation}
for \qsq\ $>$ 8 \gevsq\ and $40 < \W < 140$ \gevcsq.

The cross section for \jpsi\ meson electroproduction, taking into account the
$\jpsi \rightarrow$ 2 leptons branching fraction 0.12 \cite{PDG}, is
\begin{equation}
\sigma(e\ p \rightarrow e\ \jpsi\ p) = 100 \pm 20\ (stat.) \pm 20\ (syst.)\ {\rm pb},
   \label{eq:xjpsi}
\end{equation}
for \qsq\ $>$ 8 \gevsq\ and $30 < \W < 150$ \gevcsq.

The ratio \jpsi\ : \rh\ is $0.64 \pm 0.13$ for \qsq\ = 10 \gevsq\
and $1.3 \pm 0.5$ for \qsq\ = 20 \gevsq.

\begin{table}[htbp]
\begin{center}
\begin{tabular}{|l|l|l|}
\hline
\multicolumn{3}{|l|}{$e \ p \rightarrow e \ \rh \ p$}  \\                 
\hline
\hline
\                                    & \multicolumn{2}{|c|}{$8 < \qsq < 12$ \gevsq}              \\
\                                    &  $40 < W < 80$ \gevcsq       & $80 < W < 140$ \gevcsq     \\
\hline
\ number of events                   & 57                           & 47                         \\
\ total correction factor            & $1.53 \pm 0.22 $             & $1.76 \pm 0.25 $           \\
\ integrated $ep$ cross section [pb] & $31.5 \pm 4.2 \pm 4.5 $      & $29.8 \pm 4.3 \pm 4.2 $    \\
\ $Q_0^2$ [\gevsq], $W_0$ [\gevcsq]  & 10, 65                       & 10, 115                    \\
\ \gsp\ cross section [nb]           & $25.8 \pm 3.4 \pm 3.7 $      & $29.4 \pm 4.3 \pm 4.2 $    \\
\hline
\hline
\                                    & \multicolumn{2}{|c|}{$12 < \qsq < 50$ \gevsq}             \\
\                                    &  $40 < W < 80$ \gevcsq       & $80 < W < 140$ \gevcsq     \\
\hline
\hline
\ number of events                   & 39                           & 37                         \\
\ total correction factor            & $1.32 \pm 0.19 $             & $1.23 \pm 0.18 $           \\
\ integrated $ep$ cross section [pb] & $18.5 \pm 3.0 \pm 2.7 $      & $16.3 \pm 2.7 \pm 2.4 $    \\
\ $Q_0^2$ [\gevsq], $W_0$ [\gevcsq]  & 20, 65                       & 20, 115                    \\
\ \gsp\ cross section [nb]           & $5.0  \pm 0.8 \pm 0.7 $      & $5.0 \pm 0.8 \pm 0.7 $     \\
\hline
\hline
\multicolumn{3}{|l|}{$e \ p \rightarrow e \ \jpsi \ p$}  \\                   
\hline
\hline
\                                    & \multicolumn{2}{|c|}{$8 < \qsq < 40$ \gevsq}              \\
\                                    & $30 < W < 90$ \gevcsq       & $90 < W < 150$ \gevcsq      \\
\hline
\hline
\ number of events                   & 15                           & 16                         \\
\ total correction factor            & $1.50 \pm 0.27 $             & $0.92 \pm 0.17 $           \\
\ integrated $ep$ cross section [pb] & $61 \pm 18 \pm 12 $          & $40 \pm 12 \pm 8  $        \\
\ $Q_0^2$ [\gevsq], $W_0$ [\gevcsq]  & 16, 65                       & 16, 115                    \\
\ \gsp\  cross section [nb]          & $7.8 \pm 2.2 \pm 1.6 $       & $12.2 \pm 3.4 \pm 2.5 $    \\
\hline
\hline
\                                    & \multicolumn{2}{|c|}{$30 < W < 150$ \gevcsq}              \\
\                                    & $8 < \qsq < 12$ \gevsq       & $12 < \qsq < 40$ \gevsq    \\
\hline
\hline
\ number of events                   & 10                           & 21                         \\
\ total correction  factor           & $1.82 \pm 0.33 $             & $0.90 \pm 0.16 $           \\
\ integrated $ep$ cross section [pb] & $49 \pm 18 \pm 10 $          & $51 \pm 13 \pm 10 $        \\
\ $Q_0^2$ [\gevsq], $W_0$ [\gevcsq]  & 10, 88                       & 20, 88                     \\
\ \gsp\ cross section [nb]           & $17.6 \pm 6.3 \pm 3.7 $      & $6.6 \pm 1.6 \pm 1.4 $     \\

\hline
\end{tabular}
\end{center}
\caption{Numbers of events and cross sections for different \qsq\  and \W\ ranges, for \rh\ and
for \jpsi\ elastic production; the \gsp\ cross sections are given for $\qsq=Q_0^2, W=W_0$.}
\label{tab:xsect}
\end{table}

\subsection{Momentum transfer distributions}
\label{sect:t}
%

\begin{figure}[htbp]
\vspace{-1.cm}
\begin{center}
\epsfig{file=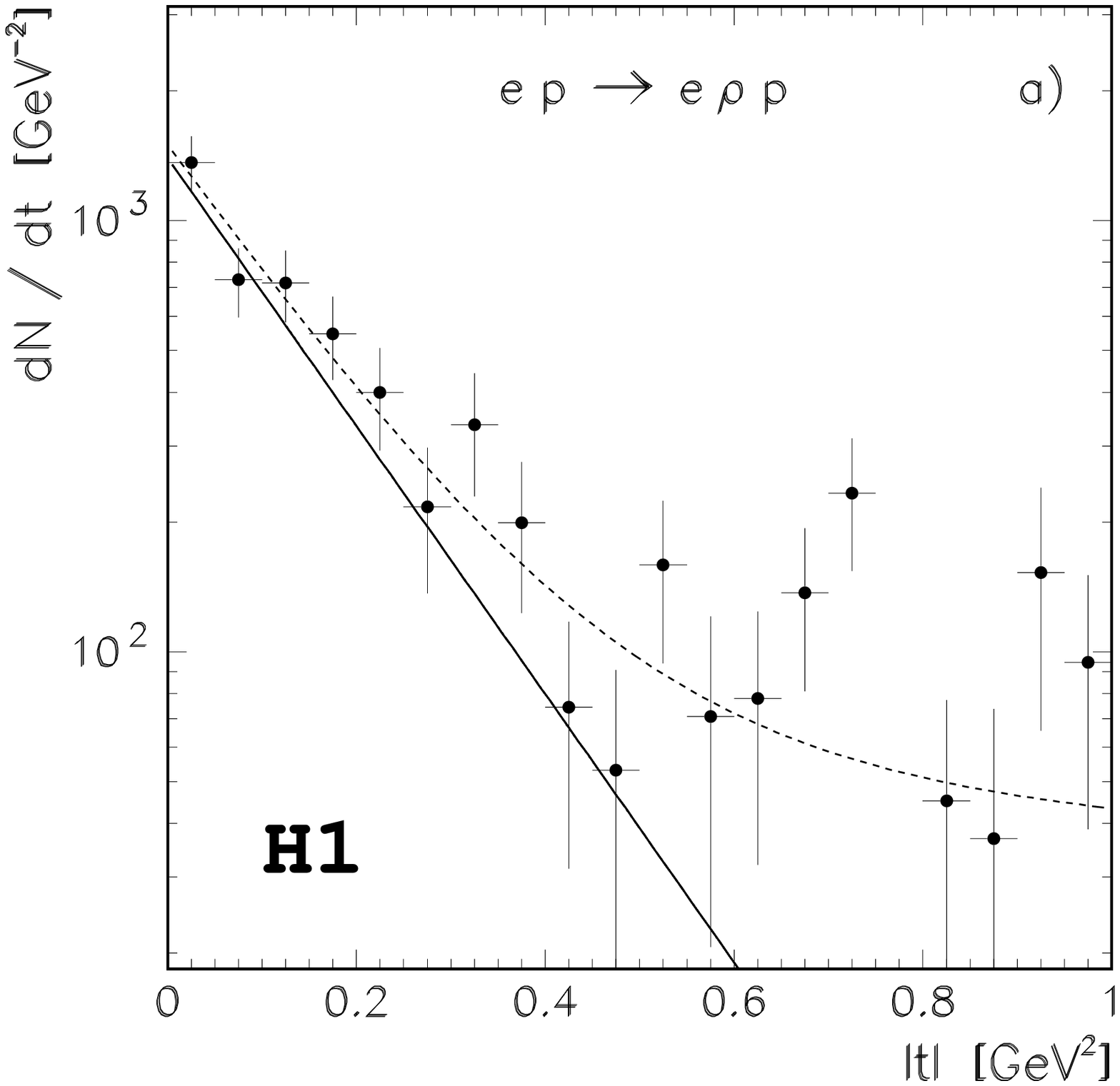,width=8cm,height=8cm}\epsfig{file=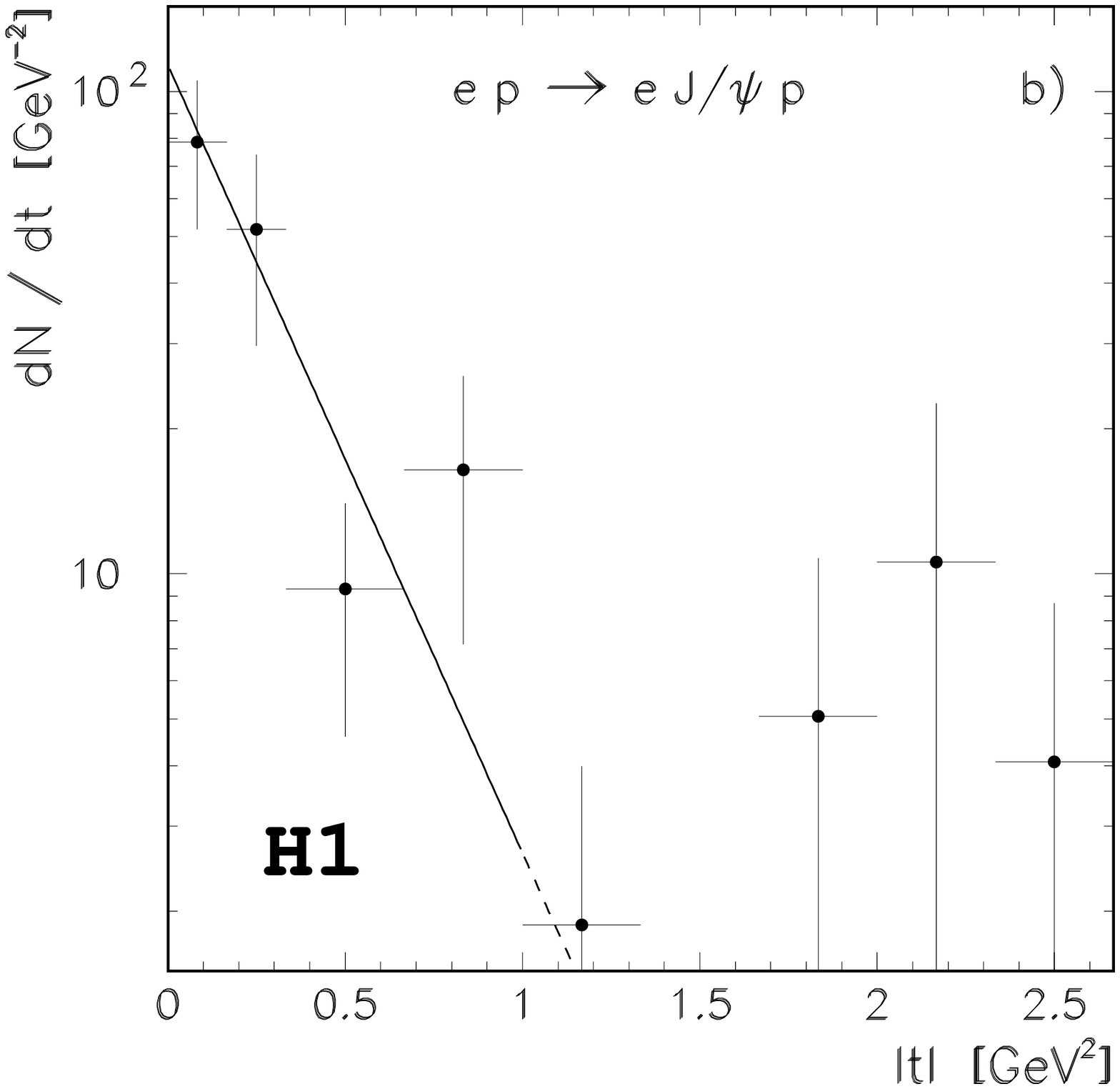,width=8cm,height=8cm}
\end{center}
\vspace{-1.cm}
\caption{\modt\ distributions for a) the \rh\ sample; the data points are not corrected
for background; the dashed line is the result of a fit taking the background into account, 
as described in the text;
b) the \jpsi\ sample, corrected for the presence of background.
Both distributions are corrected for acceptance, losses and smearing effects.
The solid lines correspond to the elastic exponential slopes.
}
\label{fig:modt_1}
\end{figure}
%

The \ttr\ distributions of the selected \rh\ and \jpsi\ events 
(Fig. \ref{fig:modt_1}) show the forward exponential peaking $\propto e^{bt}$ 
characteristic of elastic interactions. 

For the \rh\ sample, the slope \bslope\ of the \ttr\ distribution is computed taking 
into account the contributions of the non-resonant and proton dissociation 
backgrounds estimated in sections \ref{sect:mass} and \ref{sect:sim}. 
Their exponential slopes were taken to be respectively $0.15 \pm 0.10$ and 
$2.5 \pm 1.0$ \gevsqm, which is consistent with the \ttr\ dependence of event
samples which approximate these contributions. 
The results quoted below are rather insensitive to the choice of these slopes. 
The \ttr\ distribution is corrected for detector effects, including the loss of elastic 
events tagged by the forward detectors. 

A fit for \modt\ $<$ 0.6 \gevsq\ gives for the elastic slope the value 
\bslope\ = $7.0 \pm 0.8 \pm 0.4 \pm 0.5$ \gevsqm\ ($\chi^2 = 9.6 \ / 10$ d.o.f.)
for \qsq\ $>$ 8 \gevsq\ and $40 < W < 140$ \gevcsq. 
The first error corresponds to the statistical precision of the fit. 
The second describes the spread of the fits according to the choice of the 
\modt\ range (0.4 to 0.6 \gevsq) and of the \eclmax\ cut value.
The third comes from the uncertainties in the sizes and shapes of the backgrounds; 
it is dominated by the error on the total contribution of the non-resonant background.
The \ttr\ slope measured by the ZEUS Collaboration is 
\bslope~= $5.1~ _{-0.9} ^{+1.2} \pm 1.0$ \gevsqm\ \cite{ZEUS}.
The values of the slopes for two \qsq\ and two \W\ domains are given in Table~\ref{tab:slope}.

\begin{table}[htbp]
\begin{center}
\begin{tabular}{|c|c|}
\hline
\hline
\multicolumn{2}{|c|}{$40 < W < 140$ \gevcsq}                                                      \\
\ $8 < \qsq < 12$ \gevsq                         & $12 < \qsq < 50$ \gevsq                        \\
\hline
\ \bslope\ = $7.8 \pm 1.0 \pm 0.7$ \gevsqm     &  \bslope\ = $5.7 \pm 1.3 \pm 0.7$ \gevsqm        \\
\hline
\hline 
\multicolumn{2}{|c|}{$8 < \qsq < 50$ \gevsq}                                                      \\
\ $40 < W < 80$ \gevcsq                          &    $80 < W < 140$ \gevcsq                      \\
\hline
\ \bslope\ = $6.2 \pm 1.0 \pm 0.7$ \gevsqm     &  \bslope\ = $8.0 \pm 1.3 \pm 0.7$ \gevsqm        \\
\hline
\hline
\end{tabular}
\end{center}
\caption{Slopes of the \rh\ meson \ttr\ distributions for different \qsq\ and \W\ domains.}
\label{tab:slope}
\end{table}

For the \jpsi\ sample, the slope value for \modt\ $<$ 1.0 \gevsq\ is 
\bslope\ = $3.8 \pm 1.2 \ _{-1.6}^{+2.0}$ \gevsqm, after subtraction of the proton 
dissociation and non-resonant backgrounds with slopes \bslope~=~2~\gevsqm. 
The systematic error is estimated by varying the background contributions by one 
standard deviation and their slopes between 0 and 3 \gevsqm. 
The combined value of the HERA experiments [17a-b] for the slope in 
photoproduction is \bslope\ = $4.0 \pm 1.0 $ \gevsqm.
Three events have \modt\ $>$ 1.1 \gevsq\ (see Fig.~\ref{fig:modt_1}b), of which one has
\eclmax\ $>$ 1 \gev. These 3 events contribute 15\% to the cross section quoted in this
paper. The $\psi^\prime$ candidate event (see section~\ref{sect:mass}) has \modt\ 
$= 0.16$ \gevsq, \ttr\ being computed including the neutral clusters attributed
to $\pi ^0$ mesons (\qsq\ = 26.3 \gevsq, \W\ = 72.6 \gevcsq).

%
%
\begin{figure}[htbp]
\vspace{-1.cm}
\begin{center}
\epsfig{file=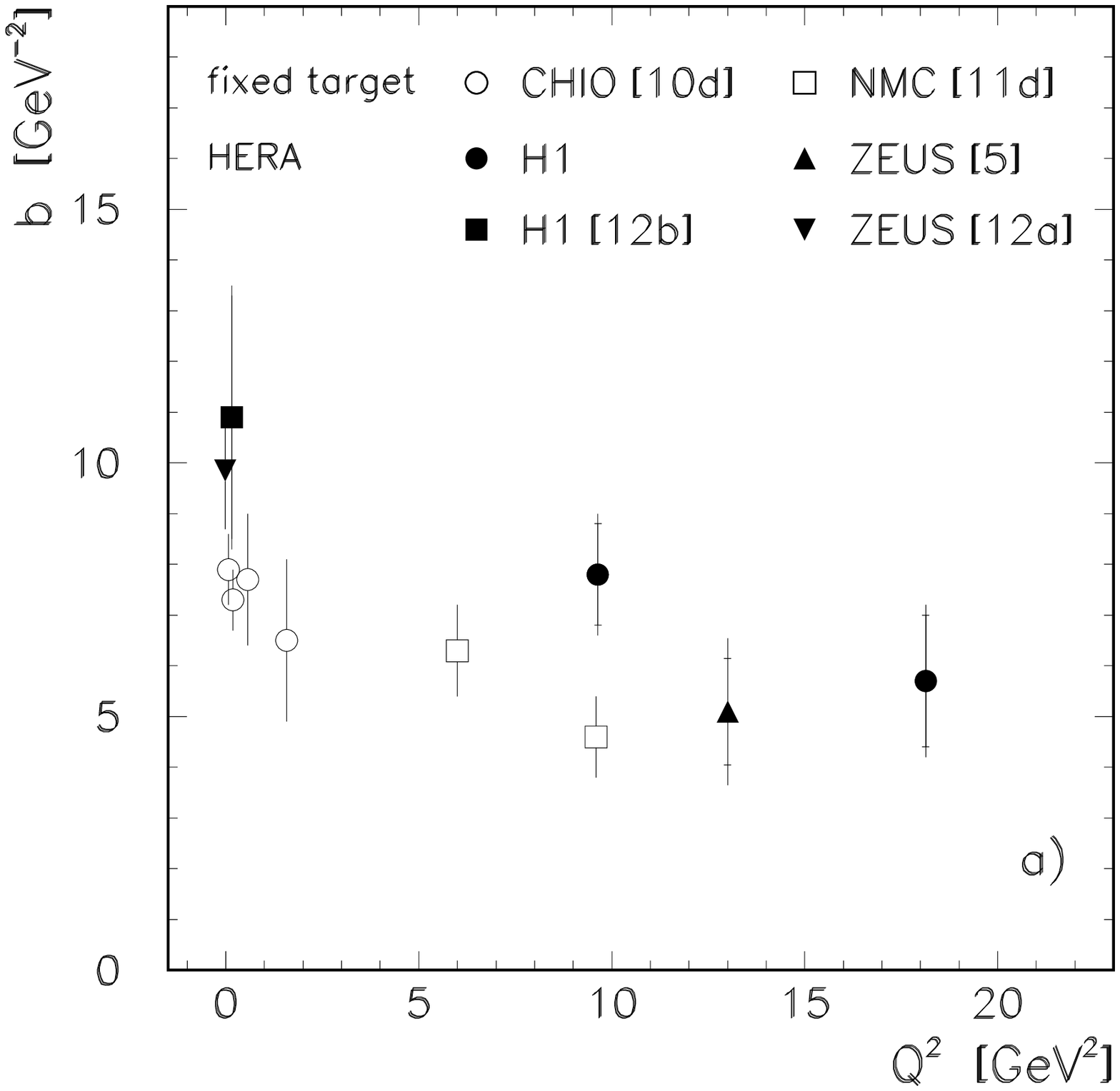,width=8cm,height=8cm}\epsfig{file=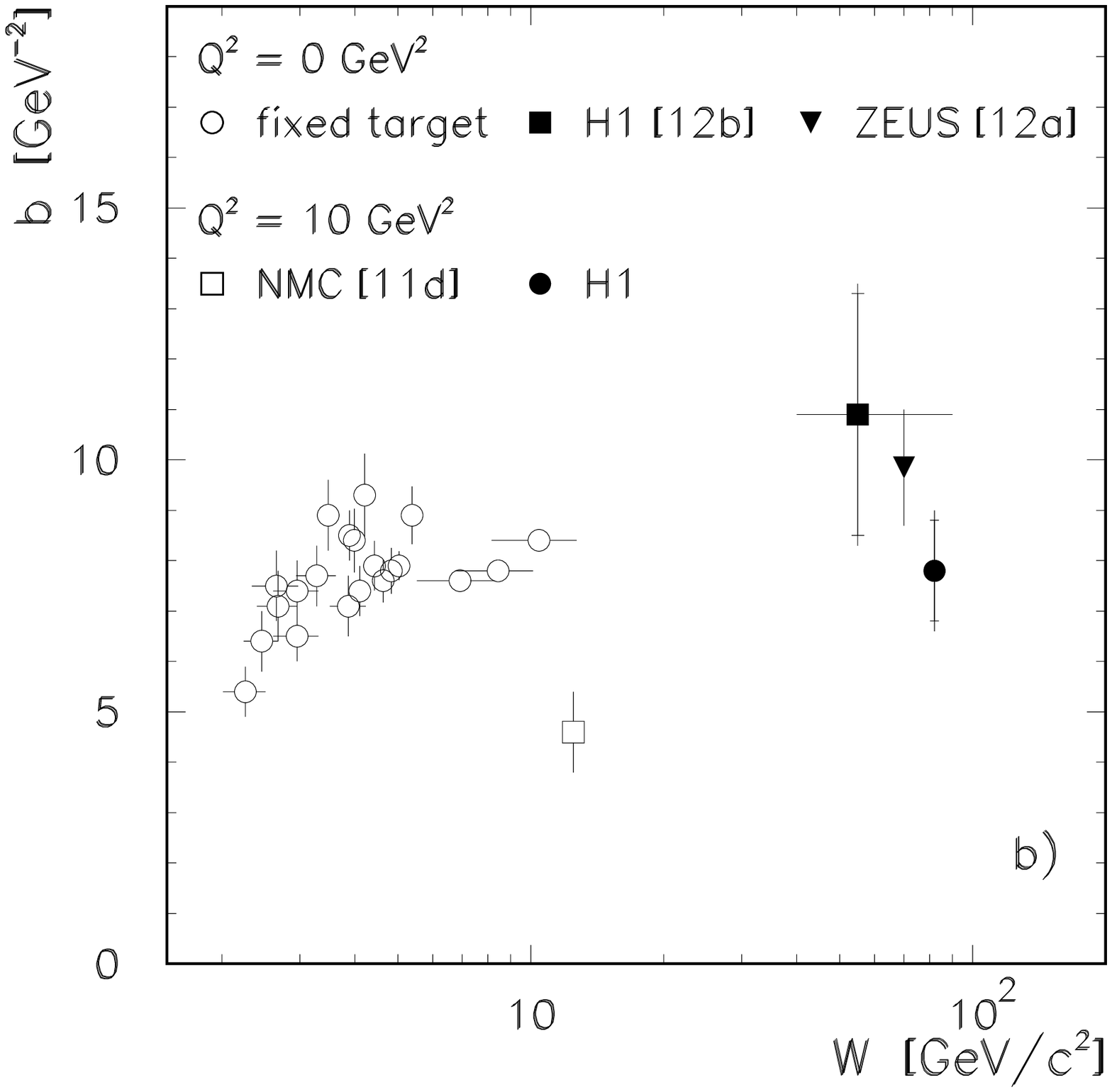,width=8cm,height=8cm}
\end{center}
\vspace{-1.cm}
\caption{ \ttr\ slope for elastic \rh\ production in fixed target and HERA
experiments a) as a function of \qsq;
b) as a function of \W\ for photoproduction and for \qsq\ $\simeq$ 10 \gevsq.}
\label{fig:modt_2}
\end{figure}
%

Fig. \ref{fig:modt_2}a shows that for the \W\ domain of the HERA experiments the decrease 
with rising \qsq\ of the \ttr\ slope for \rh\ elastic production is similar to that 
observed at lower \W.

The comparison of the NMC and H1 results for \qsq\ $\simeq$ 10 \gevsq\ (Fig.~\ref{fig:modt_2}b) 
shows an increase of the \ttr\ slope with energy. 
This shrinkage of the elastic peak with \W\ (or $\sqrt {\s}$) is observed in diffractive 
hadron interactions \cite{Goulianos} and in photoproduction (see the comparison of 
fixed target and HERA results in Fig. \ref{fig:modt_2}b). 
In the framework of Regge theory, for pomeron exchange and in terms of the exponential 
parameterisation, the shrinkage of the elastic peak can be written
\begin{equation}
b(W^2) = b(W^2=W_0^2) + 2 \ \alpha^\prime  \ \ln(W^2 / W_0^2),
                                              \label{eq:b}
\end{equation}
where $\alpha^\prime$ is the slope of the effective pomeron Regge trajectory:
\begin{equation}
\alpha_{\PO}(t) =  \alpha_{\PO}(0)+ \alpha^\prime  \ \ttr.   
                                              \label{eq:pom}
\end{equation}

Applying relation (\ref{eq:b}) to the \qsq\ = 10 \gevsq\ results (with  
statistical and systematic errors combined quadratically)
gives for $\alpha^\prime$ the value 0.41 $\pm$ 0.18 \gevsqm, in agreement with 
a value of 0.25 \gevsqm\ deduced from hadronic interactions \cite{D_L_alphaprim}. 
For the H1 data alone, there is also an indication for an increase of the slope 
with \W\ (see Table~\ref{tab:slope}).

\subsection{\qsq\ dependence of the cross sections}
\label{sect:q2}

\begin{figure}[htbp]
\vspace{-1.cm}
\begin{center}
\epsfig{file=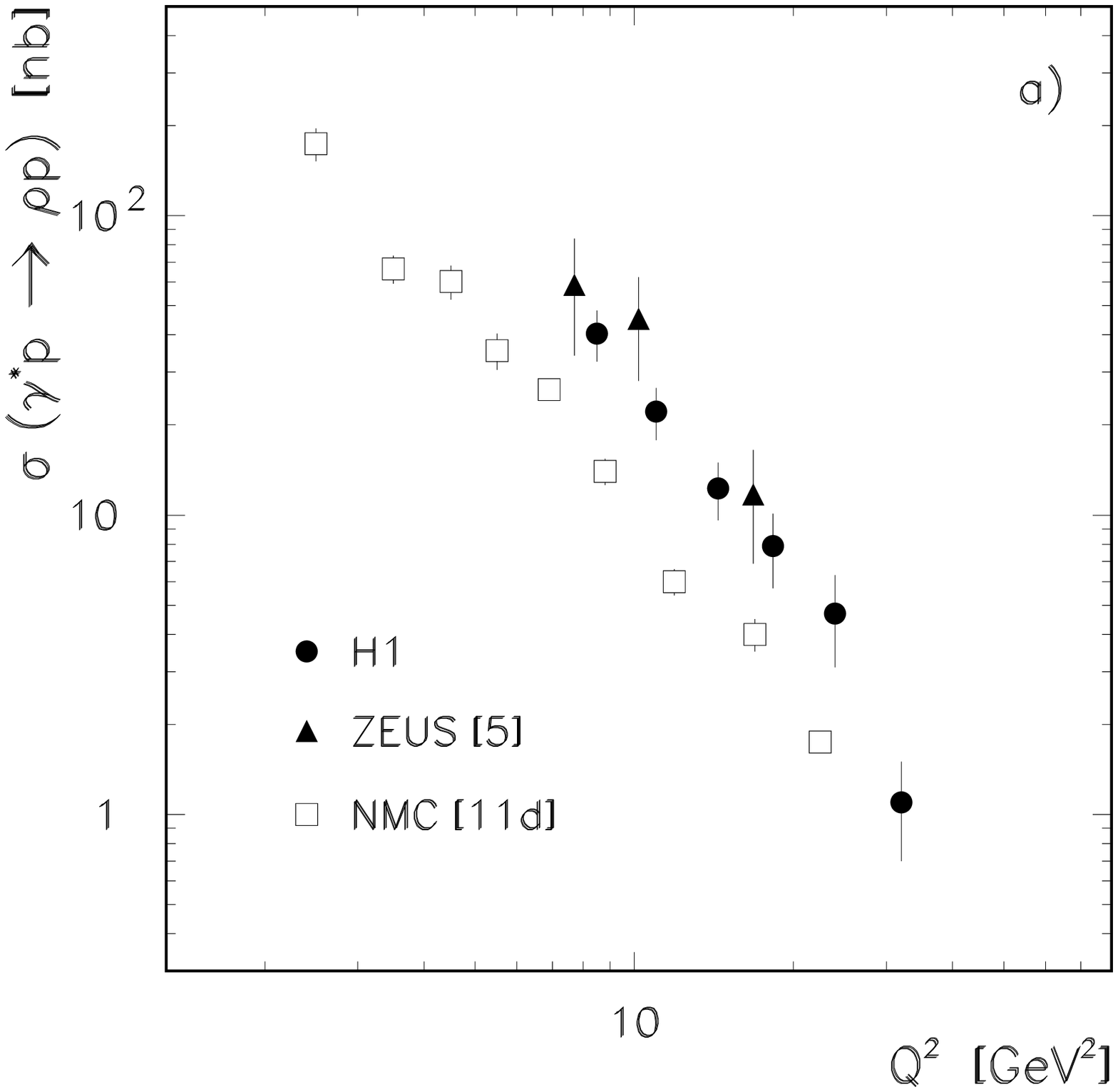,width=8cm,height=8cm}\epsfig{file=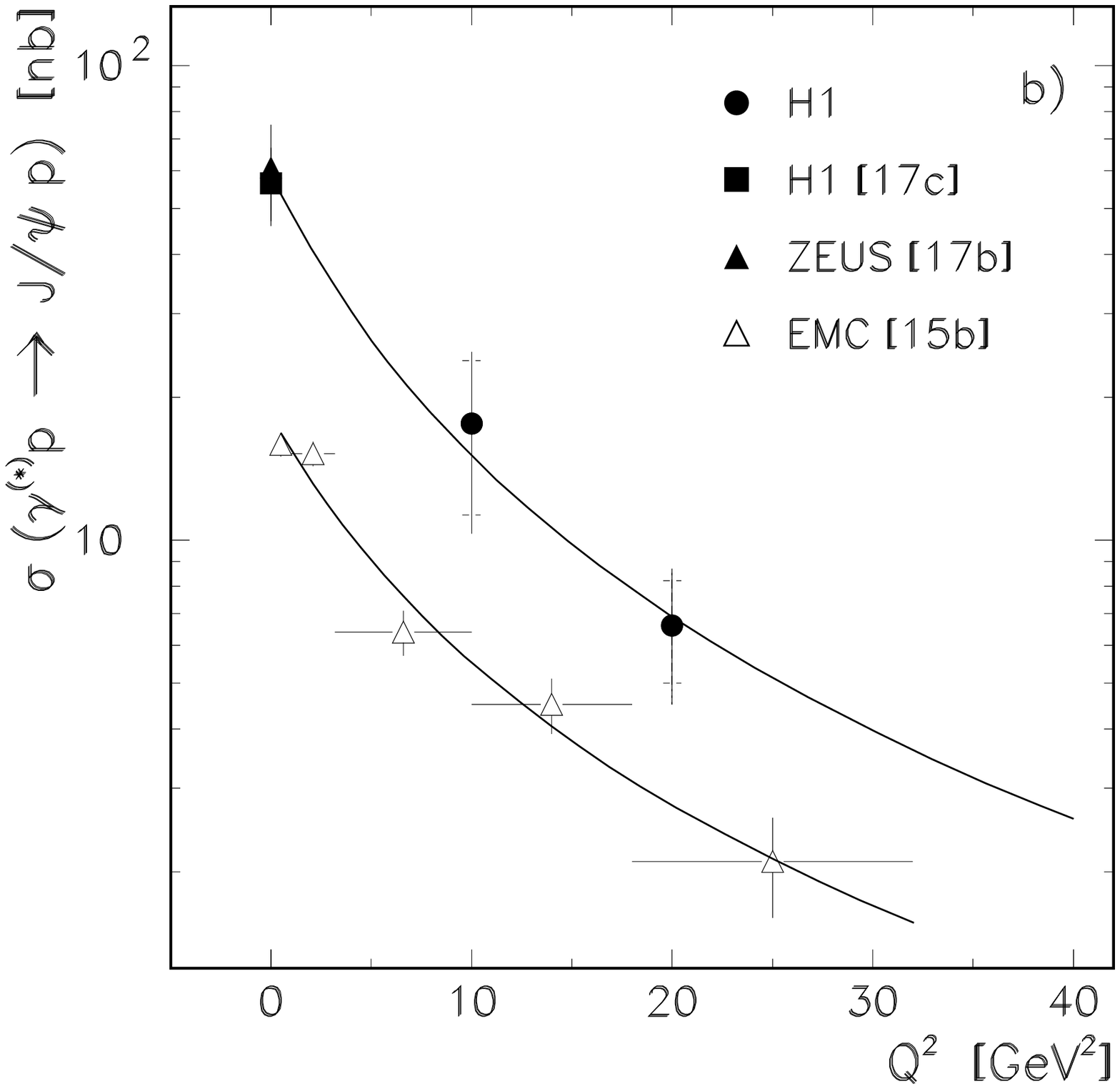,width=8cm,height=8cm}
\end{center}
\vspace{-1.cm}
\caption{ 
\qsq\ dependence of the $\gamma ^{(*)}p \rightarrow V p$ cross section
a) for \rh\ production, the ZEUS points being given for the restricted \x\ range 
$0.0014 < x < 0.004$; b) for \jpsi\ production$^6$, the curves being the 
result of the fits described in the text.}
\label{fig:q2}
\end{figure}

The \qsq\ dependence of the total \gsp\ cross section 
($\sigma_{tot} = \sigma_T + \varepsilon \ \sigma_L$) for the elastic \rh\ meson 
production by virtual photons (Fig. \ref{fig:q2}a) can be described by $Q^{-2n}$ 
with $n = 2.5 \pm 0.5 \pm 0.2$. 
In extracting the dependence on \qsq\ of the cross section, correction has been
made for the presence of non-resonant background,
for which $n = 1.5 \pm 0.2$ as obtained from the events with \modt\ $>$ 0.5 \gevsq\
or \eclmax\ $>$ 1 \gev.
The second error on $n$ reflects the uncertainty on the background size and shape and 
the spread of the results according to the details of the fitting procedure.
The cross section dependence for the present data is close to that obtained by NMC 
($n = 2.02 \pm 0.07$) and by ZEUS ($n = 2.1 \pm 0.4\ ^{+0.7}_{-0.3}$ for
$0.0014 < x < 0.004$).
It should be noted, however, that the NMC data span a large range in the 
polarisation parameter $\varepsilon$
from $\varepsilon = 0.50$ at \qsq\ =  2.5 \gevsq\  to $\varepsilon = 0.80$ 
for \qsq\ $>$ 10 \gevsq, whereas the HERA data are for $\varepsilon = 0.99$. 
Although the \qsq\ dependence is probably sensitive to this kinematical effect, it
was not taken into account because the evolution of \R\ with \qsq\ in the NMC
data (see eq. \ref{eq:sigma}) is not published.
The differences in the absolute normalisations are discussed in section \ref{sect:w}.
%
%
%

The \qsq\ dependence of the \jpsi\ production cross section at HERA is shown in 
Fig.~\ref{fig:q2}b.
The errors on the high \qsq\ data points include the uncertainty in the
\qsq\ dependence of the background.  
The evolution from photoproduction to high \qsq\ is well described by   
$1 \ / \ (\qsq + m_\psi^2)^n$ with $n = 1.9 \pm 0.3\ ({\it stat.})$. 
This is similar to the \qsq\ dependence of the low energy EMC results \cite{EMC_jpsi}, 
for which a fit of the data shown in Fig.~\ref{fig:q2}b gives $n = 1.7 \pm 0.1$.

\subsection{\W\ dependence of the cross sections}
\label{sect:w}

\begin{figure}[htbp]
\vspace{-1.5cm}
\begin{center}
\epsfig{file=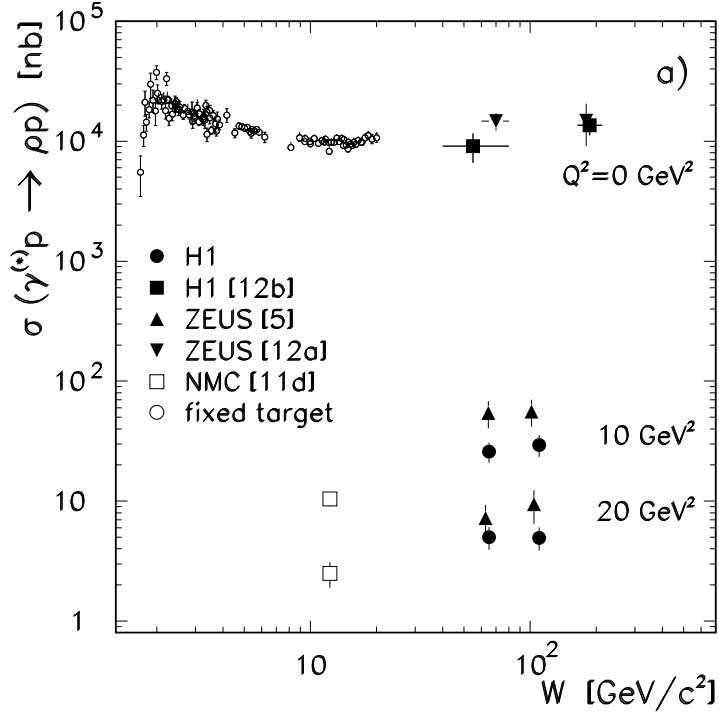,width=10.5cm,height=10.5cm}
\end{center}
\vspace{-1.cm}
\begin{center}
\epsfig{file=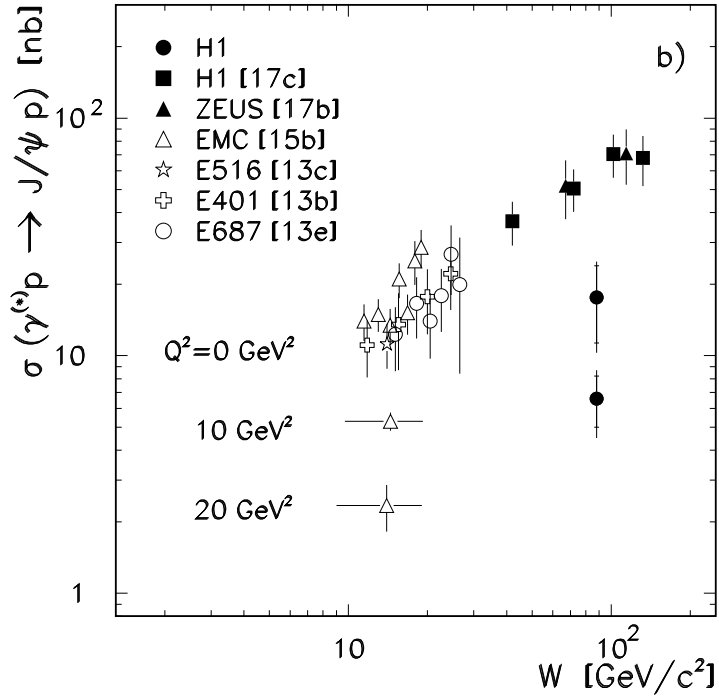,width=10.5cm,height=10.5cm}
\end{center}
\caption{\W\ dependence of the $\gamma ^{(*)} \ p \rightarrow \ V \ p$ cross section in fixed
target and HERA experiments a) for \rh\ production (computed for \mpipi\ $<$ 1.5 \gevcsq); 
b) for \jpsi\ production. For the
\rh\ data, an overall normalisation uncertainty of 31\% for ZEUS and of 20\% for NMC 
is not included in the plot.}
\label{fig:w}
\end{figure}

The \W\ dependence of the \rh\ (for \mpipi\ $<$ 1.5 \gevcsq) and \jpsi\ production 
cross sections is shown on Fig. \ref{fig:w} for \qsq\ $=$ 10 and 20 \gevsq. These values 
are chosen in order to minimise the bin centre corrections for the \rh\ analysis. 

%

The ZEUS results in Fig. \ref{fig:w}a have been scaled to the \qsq\ values of the H1
measurements using the dependence of the latter (see section \ref{sect:q2}).
These results can be directly compared to the H1 results although they include no 
explicit cut-off on the \mpipi\ mass. 
Indeed, the cross sections quoted in \cite{ZEUS} are determined assuming a non-relativistic 
Breit-Wigner mass distribution with mass independent width, integrated over the full 
kinematical range.
It turns out that this procedure leads to a cross section closely similar to that 
obtained using the relativistic form of eq. (\ref{eq:b_w}-\ref{eq:GJ}) for 
\mpipi\ $<$ 1.5 \gevcsq, as is done in the present experiment.
The ZEUS results are higher than those of H1, but the discrepancy is  
not very significant when the overall 31\% systematic error
on the ZEUS results \cite{ZEUS}, which is not included in the plot, is taken into account.
Other differences between the results of the two experiments are observed: 
both the ZEUS \ttr\ (section \ref{sect:t}) and $\cos\theta^*$ distributions 
(section \ref{sect:cos}) are flatter than for H1. 
It is worth emphasizing in this context that the event by event selection in the H1
analysis, which uses the forward detectors, provides a very clean elastic \rh\ sample.

The NMC results shown in Fig.~\ref{fig:w}a are obtained from the published 
\qsq\ dependence of the cross section for interactions on deuterium (Fig. 4 of [11d]), 
which has small nuclear corrections \cite{Sandacz,Nuov_cim}.  
They are corrected for the different values of the polarisation parameter $\varepsilon$ in 
the two experiments (the NMC measurement of \R\ is used).
The published NMC cross section was computed using for the \rh\ resonance the Breit-Wigner
parameterisation given by eq. (\ref{eq:b_w}) and (\ref{eq:Gmm}), for \mpipi\ $<$ 1.5
\gevcsq\ \cite{Sandacz}. 
The results of the two experiments can thus be directly compared.
An overall 20\% systematic uncertainty [11d] is not included in the plot.

A significant increase with energy of the elastic \gsp\ cross section is observed 
from the NMC to the HERA domains. 
Following section \ref{sect:models}, it is parameterised
as $d\sigma / dt \ (t=0)\ \propto \W^{4 \delta}$.
The fitted values of $\delta$ are for \rh\ elastic production:
\begin{equation}
\qsq = 10\ \gevsq\ :\ \  \delta = 0.14 \pm 0.05,
                                       \label{eq:del1}
\end{equation}
\begin{equation}
\qsq = 20\ \gevsq\ : \ \ \delta = 0.10 \pm 0.06.
                                       \label{eq:del2}
\end{equation}
The errors result from the combination of statistical and systematic errors of both 
experiments, including the 20\% normalisation uncertainty for NMC.

The measurements (\ref{eq:del1}) and (\ref{eq:del2}) take into account the following 
effects:

$-$ the $d\sigma / dt \ (t=0)$  cross sections are obtained by multiplying the total cross
sections by the corresponding \bslope\ slopes. 
The H1 slopes given in Table \ref{tab:slope} were used 
and the corresponding NMC slopes were computed\footnote{\ For $\qsq = 10$ \gevsq, 
the use of the measured NMC slope instead would lead to an additional increase of 
$\delta$ by 0.03. No measurement of the NMC slope is published for 
$\qsq = 20$~\gevsq.}
according to the shrinkage description given by eq. (\ref{eq:b}) with 
$\alpha^\prime = 0.25$ \gevsqm.

$-$ the cross section definition\footnote{\ See eq. (23.32) 
and (23.36), p. 1292 of \cite{PDG}.} contains a kinematical factor due to phase space 
integration, involving the centre of mass energy \W, the mass squared of the particles 
and \qsq.
This factor is not part of the study of the interaction dynamics contained
in the \W\ evolution of the matrix element. 
As the \qsq\ values considered here are rather large compared to $W^2$ for the NMC 
experiment and small for H1, there is a rising contribution to the \W\ dependence
amounting to 12\% (26\%) between NMC and H1 energies for $\qsq = 10$ (20) \gevsq.
This corresponds to a decrease of $\delta$ by 0.02 (0.03), which is included in the 
measurements (\ref{eq:del1}) and (\ref{eq:del2}).

Additional effects may have to be taken into account:

$-$ model predictions are often computed for the longitudinal cross 
section $\sigma_L$ and not for the total cross section 
$\sigma_{T} + \varepsilon \ \sigma_L$.
Taking into account the difference in the \R\ values for the two experiments
(see section \ref{sect:cos}), the $\delta$ values for $\sigma_L$ alone would be increased by 
0.02 with respect to the values (\ref{eq:del1}) and (\ref{eq:del2}) 
(a possible \qsq\ dependence of \R\ is not considered).

$-$ at the NMC energies, reggeon exchange could contribute significantly to elastic 
\rh\ production. 
Following the parameterisation obtained by Donnachie and Landshoff for the forward 
amplitude (eq. (9) of \cite{DL_1995}) and assuming it holds for high \qsq, the 
contributions to the $d\sigma / dt \ (t=0)$ cross sections of the purely reggeon 
exchange and of the reggeon$-$pomeron interference term are, respectively, 4\% and 
28\% of the pomeron exchange contribution (0\% and 4\% at HERA).
To extract the forward differential cross sections from the measured total cross
sections, assumptions have to be made concerning the relevant \bslope\ slopes.
For the purely reggeon exchange term, the value 0.83 is used for the slope of the 
Regge trajectory \cite{Cudell_Kang}, with the same parameter $b(W_0^2) = 2.5$
\gevsqm\ as for the pomeron ($W_0 = 1$ \gevcsq). 
The slope for the interference term is chosen as the average of the pomeron and the
reggeon slopes. 
When these contributions are subtracted, the value of $\delta$ is increased by 0.02.

Not including an error for theoretical uncertainties, the values of $\delta$ for $\sigma_L$ 
and pomeron exchange only are
$ 0.18 \pm 0.05\ {\rm for} \ \qsq = 10\ \gevsq$ and 
$ 0.14 \pm 0.06\ {\rm for} \ \qsq = 20\ \gevsq.$
%
%

%

The \W\ dependence of \jpsi\ production is presented in Fig.~\ref{fig:w}b\footnote{\ 
All results presented in Fig. \ref{fig:w}b have been rescaled to take into account the 
latest measurements of the \jpsi\ branching fractions:
$B(\jpsi \rightarrow e^+e^-) = 5.99 \pm 0.25 \%, 
B(\jpsi \rightarrow \mu^+\mu^-) = 5.97 \pm 0.25 \%$ \cite{PDG}.}
for \qsq\ $\simeq$ 0, 10 and 20 \gevsq. 
A steep increase of the photoproduction cross section is observed from low energy 
to the HERA experiments.
For higher \qsq\ values, a similar increase is observed between the EMC and the H1
measurements. 
However, quantitative comparisons should be taken with caution in view of normalisation 
uncertainties and the possible presence of inelastic background in the fixed target data.

\subsection{\rh\ decay angular distribution}
\label{sect:cos}

\begin{figure}[htbp]
\vspace{-1.cm}
\begin{center}
\epsfig{file=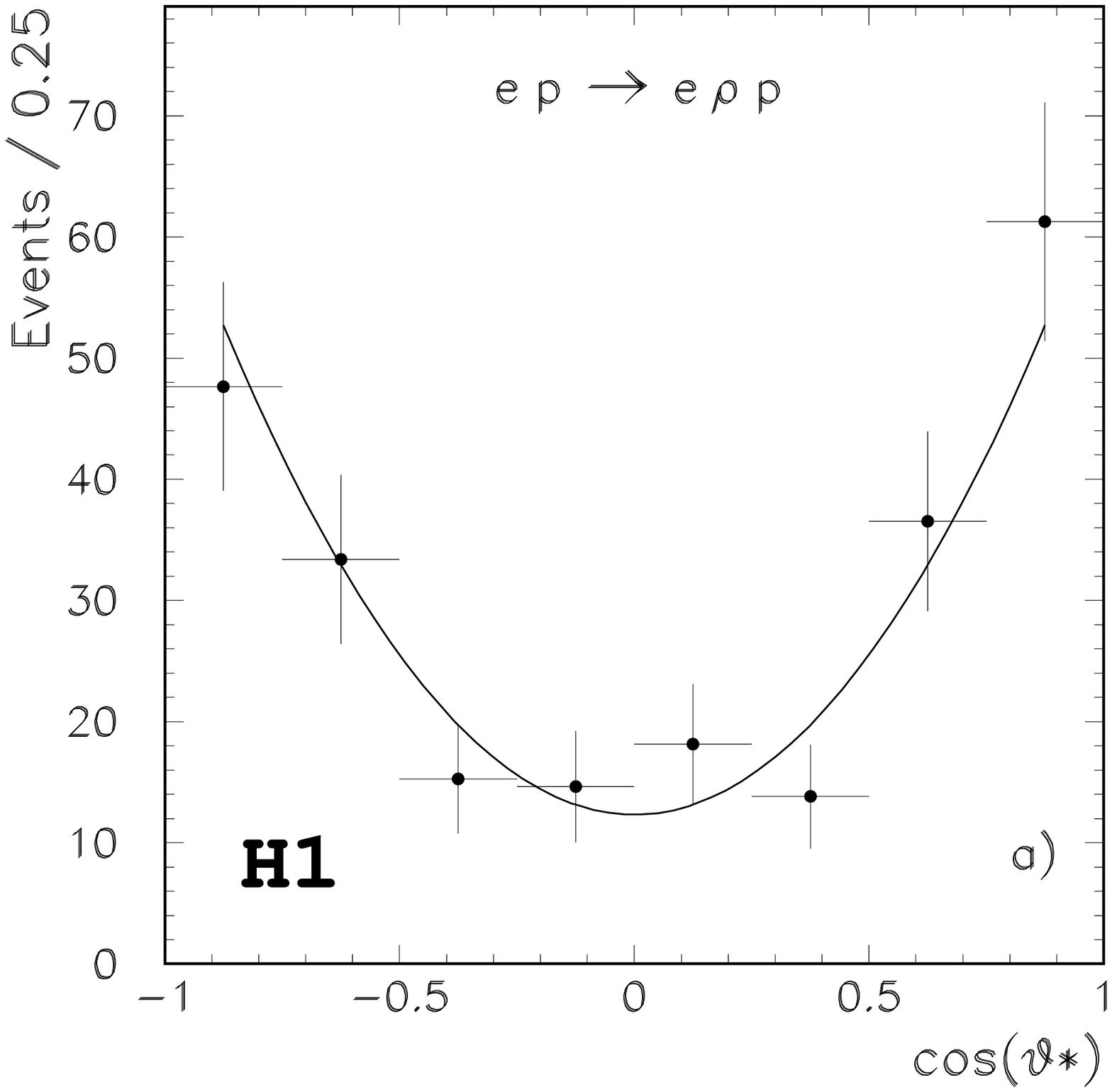,width=8cm,height=8cm}\epsfig
{file=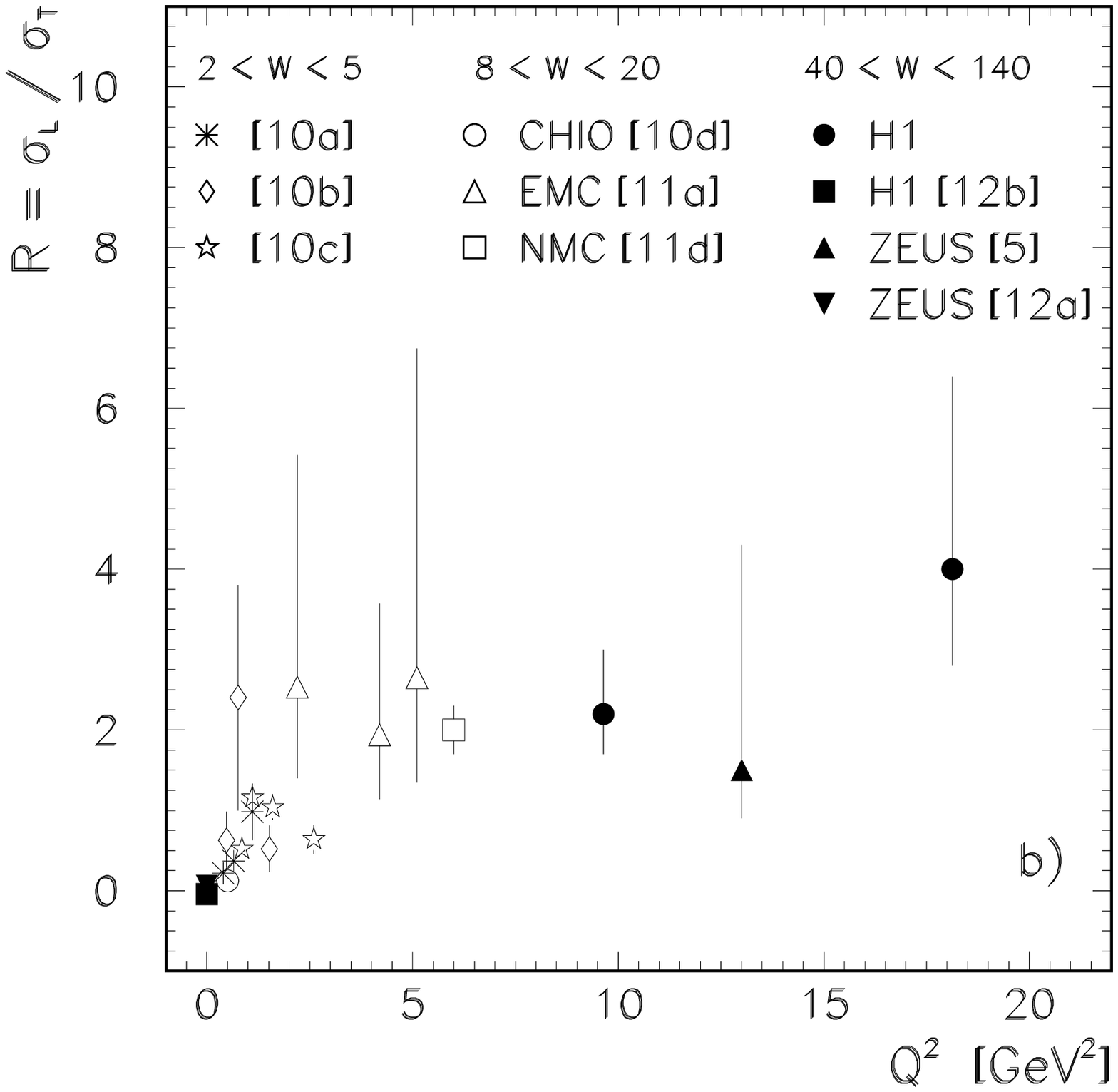,width=8cm,height=8cm}
\end{center}
\vspace{-1.cm}
\caption{a) \costhst\ distribution for the \rh\ events, the superimposed curve being 
the result of a fit of eq. (\ref{eq:costhst}) to this distribution; b) \qsq\ dependence
of $\R = \sigma_L \ / \sigma_T$ for \rh\ elastic production, from several experiments.}
\label{fig:costhst}
\end{figure}

The acceptance corrected \costhst\ distribution for the selected \rh\ sample is shown
in Fig.~\ref{fig:costhst}a. 
After subtraction of the non-resonant background, which is consistent with being flat
in \costhst, and
correction for detector effects, the fit of eq. (\ref{eq:costhst}) to this distribution
gives \rzzzz\ = $0.73 \pm 0.05 \pm 0.02$. 
The first error is statistical, the second reflects the uncertainty on the background 
subtraction. 
Assuming $SCHC$, relation (\ref{eq:R}) gives 
$R = \sigma_L \ / \sigma_T = 2.7\ _{-0.5\ -0.2}^{+0.7\ +0.3}$, 
with $\av {\varepsilon}$~=~0.99.
 
%
%

The value of \R\ is shown in Fig. \ref{fig:costhst}b together with fixed target
measurements\footnote{\ EMC measurements [11a] with largest errors have been omitted.} 
and given in Table \ref{tab:R} for two values of \qsq\ and of \W. 
Compared with results at low \qsq, a clear increase of \R\ with \qsq\ is observed. 

\begin{table}[htbp]
\begin{center}
\begin{tabular}{|c|c|}
\hline
\hline
\multicolumn{2}{|c|}{$40 < W < 140$ \gevcsq}                                                    \\
\ $8 < \qsq < 12$ \gevsq                         & $12 < \qsq < 50$ \gevsq                      \\
\hline
\  $\R =  2.2\ _{-0.5}^{+0.8}$                     & $\R =  4.0\ _{-1.2}^{+2.4}$                    \\
\hline
\hline 
\multicolumn{2}{|c|}{$8 < \qsq < 50$ \gevsq}                                                    \\
\ $40 < W < 80$ \gevcsq                          & $80 < W < 140$ \gevcsq                       \\
\hline
\ $\R =  2.2\ _{-0.5}^{+0.8}$                      & $\R =  3.7\ _{-1.1}^{+1.9}$                    \\
\hline
\hline
\end{tabular}
\end{center}
\caption{Values of $R = \sigma_L \ / \sigma_T$ for \rh\ elastic production in different 
\qsq\ and \W\ domains.}
\label{tab:R}
\end{table}

An attempt was made to test the hypothesis \cite{Ginzburg} that 
the \rh\ meson should be completely longitudinally polarised for \modt\ $\gg$ 
$\Lambda_{QCD}^2$, by dividing the data with \modt\ $<$ 0.5 \gevsq\ into two samples 
with \modt\ respectively smaller and larger than 0.15 \gevsq.
The predicted effect was not observed, but the imposed cut is probably 
too low to provide a sensitive test of the prediction.


\section{Discussion and conclusions}

The production of elastic \rh\ and \jpsi\ mesons by virtual photons has been measured 
at HERA with the H1 detector. 
Samples of 180 and 31 events, respectively, have been collected with
\qsq\ $>$ 8 \gevsq\ and $40 < \W < 140$ \gevcsq\ (30 $-$ 150 \gevcsq\ for the \jpsi), 
for an integrated luminosity of 2.8 (3.1) pb$^{-1}$.
Most of the proton dissociation background is removed using the forward components 
of the H1 detector and the small residual backgrounds are corrected for.

A major interest of the study of elastic vector meson production is that predictions
have been proposed for the differential cross sections both in the framework of a
soft, non-perturbative approach, and based on perturbative QCD calculations for
hard processes. 

The main difference between the predictions of the soft and of the hard approaches
concerns the rise of the cross section with energy, expected respectively to be
slow or fast.
For \rh\ production at high \qsq, the \W\ dependence of the cross section attributed
to pomeron exchange is parameterised as 
$d\sigma / dt \ (t=0)\ \propto \W^{4 \delta}$.
Using the NMC and the present H1 results, the measured values of $\delta$ for the total 
cross section $\sigma_T + \varepsilon \ \sigma_L$ are 
$ \delta = 0.14 \pm 0.05\ {\rm for} \ \qsq = 10\ \gevsq$ and 
$ \delta = 0.10 \pm 0.06\ {\rm for} \ \qsq = 20\ \gevsq.$
For the longitudinal cross section $\sigma_L$ alone, and taking into account possible
reggeon exchange, these values would be 
$ \delta = 0.18 \pm 0.05$ and 
$ \delta = 0.14 \pm 0.06$,
respectively.

For the soft pomeron model of Donnachie-Landshoff, the expected value 
is 0.08.
For the hard approach, it is presumably in the range $0.20 - 0.25$ [3b]. 
The present measurements thus lie between the values expected for these
two types of models. 

The suggestion \cite{Brodsky} that the \rh\ cross section measurement 
may provide information on the gluon distribution in the proton is applicable
only when the hard regime is reached. 
As this condition does not seem to be fulfilled for the present \W\ and \qsq\ ranges, 
an attempt to extract the gluon distribution from these data seems premature. 

In contrast, the \jpsi\ production cross section for \qsq\ $>$ 8 \gevsq\ increases 
strongly from the fixed target to the HERA region. 
This increase is of the same order as in photoproduction. 
This indicates that a hard regime is reached for \jpsi\ production already at 
low \qsq, which could be related to the smaller spatial extent of the wave function
and the large scale provided by the charm quark mass.

A major result of the present measurement is the similarity of the cross
sections for \rh\ and \jpsi\ elastic production.
Whereas \jpsi\ photoproduction, which is suppressed by factors of 100 to 1000 
with respect to \rh, is not well described by the ``quark counting rule'', quark 
flavour symmetry appears to be approximately restored for \qsq\ of 10 to 20 \gevsq.
Such a behaviour is expected both in the soft and the hard models. 
However, this evolution is observed to be faster than for some hard models
(a ratio 1/2 has been proposed for \qsq\ $\simeq \ 100 $ \gevsq\ \cite{F_K_S}).

The \ttr\ dependence of the production differential cross sections for \rh\ and \jpsi\
mesons are found to be well described at low \ttr\ values by exponential dependences 
$e^{bt}$. 
For the \rh\ sample, \bslope\ is  $7.0 \pm 0.8\ (stat.) \pm 0.4\ (syst.) \pm 0.5\ (bg.)$ \gevsqm.
This value is smaller than for photoproduction at HERA, showing that the decrease of the 
slope with rising \qsq\ observed in fixed target experiments extends to the HERA
regime.
This behaviour can be attributed to the decrease of the $q \bar q$ transverse 
separation in the photon with rising \qsq.

The evolution of the slope with \W\ is sensitive to the interplay of soft and hard 
effects in \rh\ production.
There is an indication that the shrinkage of the elastic peak observed in hadron 
interactions and \rh\ photoproduction also occurs in the present electroproduction 
data, at large \qsq.
This is as expected from Regge predictions based on soft pomeron exchange, in contrast
with the little shrinkage predicted in perturbative calculations for hard processes 
\cite{F_K_S}.
 
For \jpsi\ production, the \ttr\ distribution is well described with 
$\bslope\ = 3.8 \pm 1.2 \ _{-1.6}^{+2.0}$
\gevsqm, which is smaller than for the \rh.
This difference can be qualitatively explained by the fact that the \jpsi\ wave function 
is more compact than the \rh\ wave function.

The \qsq\ dependence of the \rh\ total cross section can be described by a power law 
$Q^{-2n}$, with $n = 2.5 \pm 0.5 \pm 0.2$.
This distribution is slightly harder than initially expected both for non-perturbative 
two gluon exchange and for hard  QCD calculations ($\propto Q^{-6}$). It is 
compatible with predictions taking into account the \qsq\ evolution of parton 
densities and the transverse motion of the quarks in the photon \cite{F_K_S}.
It should be noted that the calculations are performed for $\sigma_{L}$
whereas the present measurement is of $\sigma_{tot}$, which includes a transverse 
contribution at the level of 20$-$25\%, with presumably a steeper 
\qsq\ dependence \cite{Kope, Brodsky}. 

The \qsq\ dependence of the \jpsi\ production cross section is parameterised 
as $1 / (Q^2 + m_\psi^2)^n$, with $n = 1.9 \pm 0.3\ (stat.)$, which is 
similar to fixed target results.

The polar angle distribution of the decay pions indicates that \rh\ mesons
are mostly longitudinally polarised: the spin density matrix element \rzzzz\ is 
$0.73 \pm 0.05 \pm 0.02$.
Assuming {\it SCHC},
$R = \sigma_L \ / \sigma_T$ is $2.7\ _{-0.5\ -0.2}^{+0.7\ +0.3}$.
The increase of \R\ with \qsq, which is predicted both by the non-perturbative model 
and the QCD calculations, is observed.
The indication in the present data of an increase of \R\ with \W\ suggests,
in the framework of hard physics, that perturbative features at high \W\ and 
moderate \qsq\ are indeed more important for longitudinal than for transverse photons.

In conclusion, the study of \rh\ and \jpsi\ meson production offers a contrasting 
picture.

For \jpsi\ mesons, hard physics effects are probably at work already for very
small \qsq. The present high \qsq\ data support this interpretation, albeit with
limited statistical precision.

For \rh\ mesons, the \W, \ttr, \qsq\ and \costhst\ dependences of the cross section 
allow a more detailed study, from which a mixed picture emerges. 
It can be speculated that the present data correspond 
to a transition regime, with interplay of hard processes, amenable to perturbative 
description, and soft processes, requiring a non-perturbative approach.
The \W\ dependence of the cross section does not provide conclusive evidence 
in favour of a purely soft or a purely hard model. 
The indication of shrinkage of the elastic peak with \W\ is important  
because it shows a continuation with increasing \qsq\ of the photoproduction behaviour 
and is at variance with expectations for a purely hard behaviour.
The observation that SU(4) flavour symmetry is restored in the \jpsi\ : \rh\ cross section 
ratio, which is a striking feature of the present measurements, is expected in both 
models.
The study of the \qsq\ and \costhst\ distributions also does not  
discriminate between the soft and hard models, since their predictions are similar.

\section*{Acknowledgements}
We are grateful to the HERA machine group whose outstanding efforts
made this experiment possible. We appreciate the immense effort of the
engineers and technicians who constructed and maintained the detector.
We thank the funding agencies for their financial support of the
experiment. We wish to thank the DESY directorate for the support
and hospitality extended to the non-DESY members of the collaboration.
We thank further J.-R. Cudell, B. Kopeliovich, A. Sandacz and 
H. Spiesberger for useful discussions.

{\Large\normalsize}

\end{document}